\documentclass[5p,times,nopreprintline]{elsarticle}

\usepackage[utf8]{inputenc}
\usepackage[T2A,T1]{fontenc}
\usepackage[russian,english]{babel}

\usepackage{graphicx,amsmath,color}

\usepackage{hyperref}

\usepackage{multirow}

\usepackage{lineno}
\usepackage{tikz}

\def\maketitle{%
 \onecolumn%
 \elsarticleprelims%
 \finalMaketitle\printFirstPageNotes
 \gdef\thefootnote{\arabic{footnote}}
 \twocolumn%
 }

\begin{document}

\begin{frontmatter}

\title{The JUNO experiment Top Tracker}

\let\affil\address

\author[6,5]{Angel Abusleme}
\author[45]{Thomas Adam}
\author[66]{Shakeel Ahmad}
\author[66]{Rizwan Ahmed}
\author[55]{Sebastiano Aiello}
\author[66]{Muhammad Akram}
\author[66]{Abid Aleem}
\author[48]{Tsagkarakis Alexandros}
\author[21]{Fengpeng An}
\author[23]{Qi An}
% ...
\affil[5]{Millennium Institute for SubAtomic Physics at the High-energy Frontier (SAPHIR), ANID, Chile}
\affil[6]{Pontificia Universidad Cat\'{o}lica de Chile, Santiago, Chile}
% ...
\affil[21]{Sun Yat-Sen University, Guangzhou, China}
% ...
\affil[23]{University of Science and Technology of China, Hefei, China}
% ...
\affil[45]{IPHC, Universit\'{e} de Strasbourg, CNRS/IN2P3, F-67037 Strasbourg, France}
% ...
\affil[48]{III. Physikalisches Institut B, RWTH Aachen University, Aachen, Germany}
% ...
\affil[55]{INFN Catania and Dipartimento di Fisica e Astronomia dell Universit\`{a} di Catania, Catania, Italy}
% ...
\affil[66]{Pakistan Institute of Nuclear Science and Technology, Islamabad, Pakistan}
% ...

\cortext[contact]{Corresponding authors}

\begin{abstract}
The main task of the Top Tracker detector of the neutrino reactor experiment Jiangmen Underground Neutrino Observatory (JUNO) is to reconstruct and extrapolate atmospheric muon tracks down to the central detector.
This muon tracker will help to evaluate the contribution of the cosmogenic background to the signal.
The Top Tracker is located above JUNO's water Cherenkov Detector and Central Detector, covering about 60\% of the surface above them.
The JUNO Top Tracker is constituted by the decommissioned OPERA experiment Target Tracker modules.
The technology used consists in walls of two planes of plastic scintillator strips, one per transverse direction.
Wavelength shifting fibres collect the light signal emitted by the scintillator strips and guide it to both ends where it is read by multianode photomultiplier tubes.
Compared to the OPERA Target Tracker, the JUNO Top Tracker uses new electronics able to cope with the high rate produced by the high rock radioactivity compared to the one in Gran Sasso underground laboratory.
This paper will present the new electronics and mechanical structure developed for the Top Tracker of JUNO
along with its expected performance based on the current detector simulation.
\end{abstract}


\end{frontmatter}

\section{Introduction} \label{introduction}

\begin{sloppypar} 
The Jiangmen Underground Neutrino Observatory (JUNO)~\cite{Djurcic:2015vqa,JUNO:2022hxd}, under installation in the South of China at a distance of about 53~km from two nuclear plants, is a multipurpose experiment with the main goal of determining the neutrino mass ordering.
For this purpose, the experiment will use a 20~kt liquid scintillator detector with a very good energy resolution.
At the same time, JUNO will also measure with a sub-percent accuracy the oscillation parameters $\theta_{12}$, $\Delta m_{21}^2$ and $\Delta m_{31}^2$.
JUNO will be able to detect neutrinos from other natural sources as, e.g., geo-neutrinos and supernova explosions~\cite{JUNO:2022hxd,JUNO:2015zny}.
\end{sloppypar}

\begin{figure}[htb]
\centering
\includegraphics[width=\linewidth]{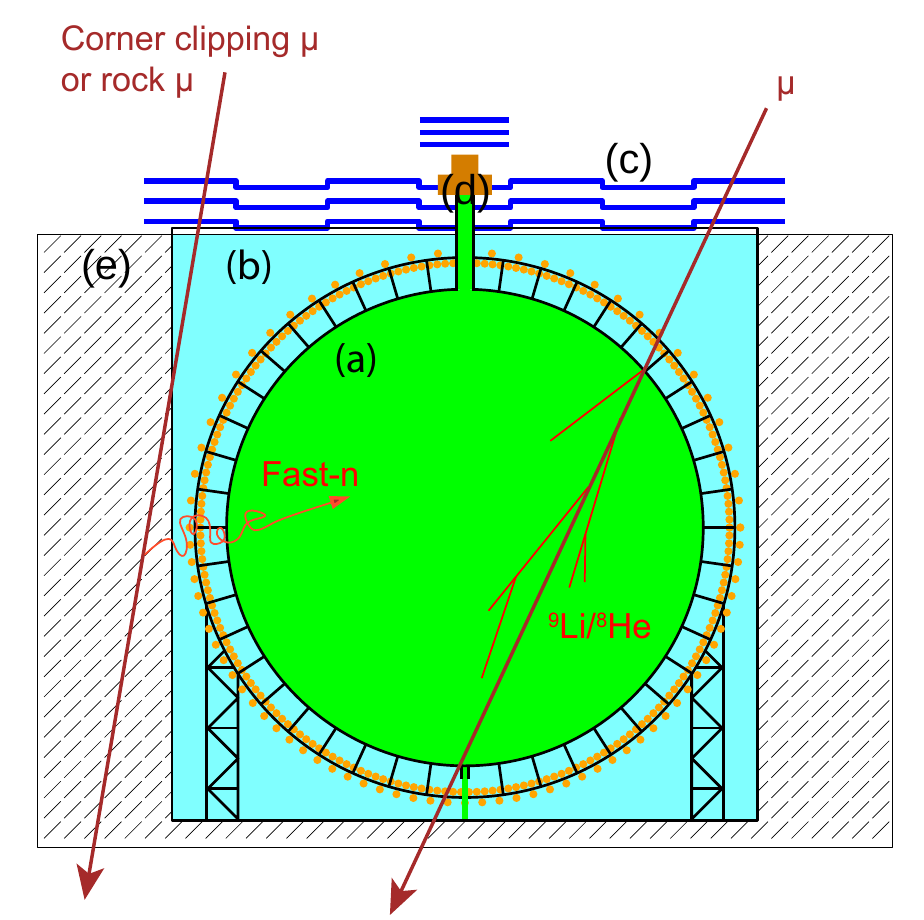}
  \caption{Schematic view of the JUNO detector with the (a) central detector, (b) water Cherenkov detector and (c) Top Tracker.
  The (d) calibration house and the chimney connecting it to the central detector are also shown.
  Configurations considered for the induced noise, from either fast neutrons or cosmogenic isotopes, in the central detector, the water Cherenkov detector and the (e) surrounding rock are also illustrated.}
\label{ttnoise}
\end{figure}

The JUNO detector is composed of a central detector containing an acrylic sphere in which the liquid scintillator is located, surrounded by 17612 large (20") and 25600 small (3") photomultipliers (PMTs).
The central detector is immersed in a water Cherenkov detector instrumented with 2400 large PMTs.
A muon tracker, called Top Tracker (TT), is placed on top of those detectors.
The water Cherenkov detector and the Top Tracker combined form the Veto system of JUNO that was conceived
to track and veto almost all muons crossing the detector.
A detailed description of the experimental equipment can be found in~\cite{JUNO:2022hxd} and the above
listed elements are also shown in Fig.~\ref{ttnoise}.
In this figure is also identified the position of the calibration house, required for the central detector
calibration, and the chimney connecting them.
In order to fit the chimney and calibration house, the central layers of the Top Tracker were moved to a higher
position, as is also shown in Fig.~\ref{ttnoise}.
The detector will be placed in an underground laboratory with an overburden corresponding to about 700~m depth (1800~m.w.e).
At this depth it is still expected to have about 15 atmospheric muons per hour and per m$^2$ with a mean energy of 207~GeV.
These muon tracks can induce a background very similar to the inverse beta decay (IBD) interactions produced by reactor neutrinos.

The most dangerous background induced by these atmospheric muons,
which are produced by the interaction of cosmic-rays with the Earth's atmosphere,
comes from the generation in the detector or in the surrounding rock
of $^9$Li and $^8$He unstable elements, and fast neutrons.
Fig.~\ref{ttnoise} presents schematically these background contributions.
The main contribution comes from $^9$Li and $^8$He produced directly in the central detector and in the water Cherenkov detector while the last one concerns the neutron production in the surrounding rock.

The present simulations show that IBD interactions and cosmogenic background mimicking these interactions have comparable rates.
For this reason the rejection of this background contribution and the study of its impact on the systematic errors is very important in JUNO.
This is the main role of the Top Tracker.
All OPERA experiment~\cite{Guler:2000bd} Target Tracker~\cite{Adam:2007ex} plastic scintillating modules have been recovered and sent to China where they will be used to form the JUNO Top Tracker.

In the OPERA experiment, the Target Tracker modules were shielded by the emulsion/lead bricks, while in the JUNO cavern the TT will be directly exposed to rock radioactivity.
On top of that, the radioactivity in the JUNO cavern is measured to be about two orders of magnitude higher than the one observed in the underground Laboratori Nazionali del Gran Sasso (LNGS) where the OPERA experiment was located.
For these reasons, the electronics of the OPERA Target Tracker had to be replaced by a significantly faster version.

The 496 OPERA Target Tracker modules are only enough to cover only about 60\% of the top surface of the JUNO central detector and thus can track only 30\% of the total muon flux.
For this reason, the TT will participate in the JUNO veto for the muons traversing it.
It is worth noticing here that about half of the atmospheric muons pass by the sides of the water Cherenkov detector without crossing the top surface where the TT is located.
The water Cherenkov detector surrounding all the JUNO central detector will be used as veto to reject events coming from all directions, including those missed by the TT.

A large number of well tagged $^9$Li and $^8$He will allow to well measure the energy spectrum and rate of this background.
The TT can also be used to select and well reconstruct a sample of muon tracks to be used to tune and validate the reconstruction algorithms based on central and water Cherenkov detectors' information.
This is possible because of the centimetre level accuracy on the position of the muon tracks crossing the TT.

This paper describes in Sec.~\ref{overview} the detection principle and TT geometry.
This is followed by a description of the expected rates in the detector along different levels of trigger
in Sec.~\ref{radioactivity}.
The Top Tracker electronics designed to perform the data readout and trigger under the expected background
conditions is extensively presented in Sec.~\ref{electronics}.
In Sec.~\ref{sec:reconstruction} the current Top Tracker reconstruction and its performance is presented.
Details on the operation modes of the Top Tracker are detailed in Sec.~\ref{operation},
which is followed by a description of a prototype of the Top Tracker in Sec.~\ref{ttproto} where these
modes were extensively tested.
Elements concerning the supporting structure for the Top Tracker modules and installation are provided in
Sec.~\ref{installation}.
Long term stability monitoring of the Top Tracker modules is presented in Sec.~\ref{sec:ageing}.
Finally, a summary and concluding remarks are presented in Sec.~\ref{sec:conclusion}.

\section{Description of the Top Tracker} \label{overview}

As described in~\cite{Adam:2007ex} presenting the OPERA Target Tracker, the TT uses 6.86~m long, 10.6~mm thick, 26.4~mm wide scintillator strips read on both sides using Wave Length Shifting (WLS) fibres and Hamamatsu\footnote{Hamamatsu Photonics K.K., Electron Tube Center, 314--5,
Shimokanzo, Toyooka-village, Iwata-gun, Shizuoka-ken,
438--0193, Japan.} $8\times8$ channels multianode photomultipliers (\mbox{MA-PMT}) H7546.
The technology used in the TT is very reliable due to the robustness of its components.
Delicate elements, like electronics and \mbox{MA-PMTs} are located outside the sensitive area where they are easily accessible.

\begin{figure}[hbt]
\centering
\includegraphics[width=\linewidth]{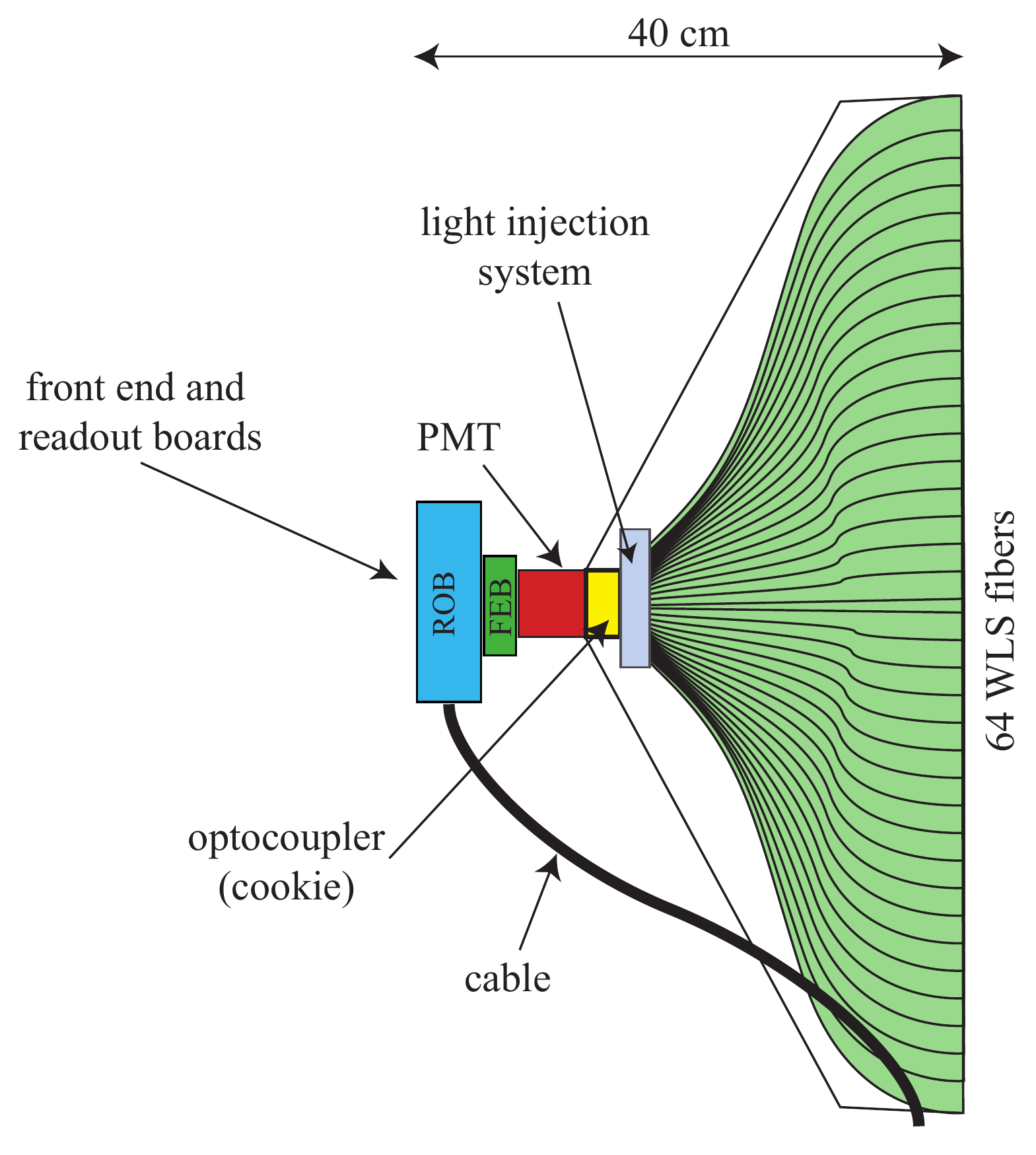}
\caption{Schematic view of an end-cap of a scintillator
strip module.}\label{endcap_schematic}
\end{figure}

\begin{figure}[hbt]
\centering
\includegraphics[width=\linewidth]{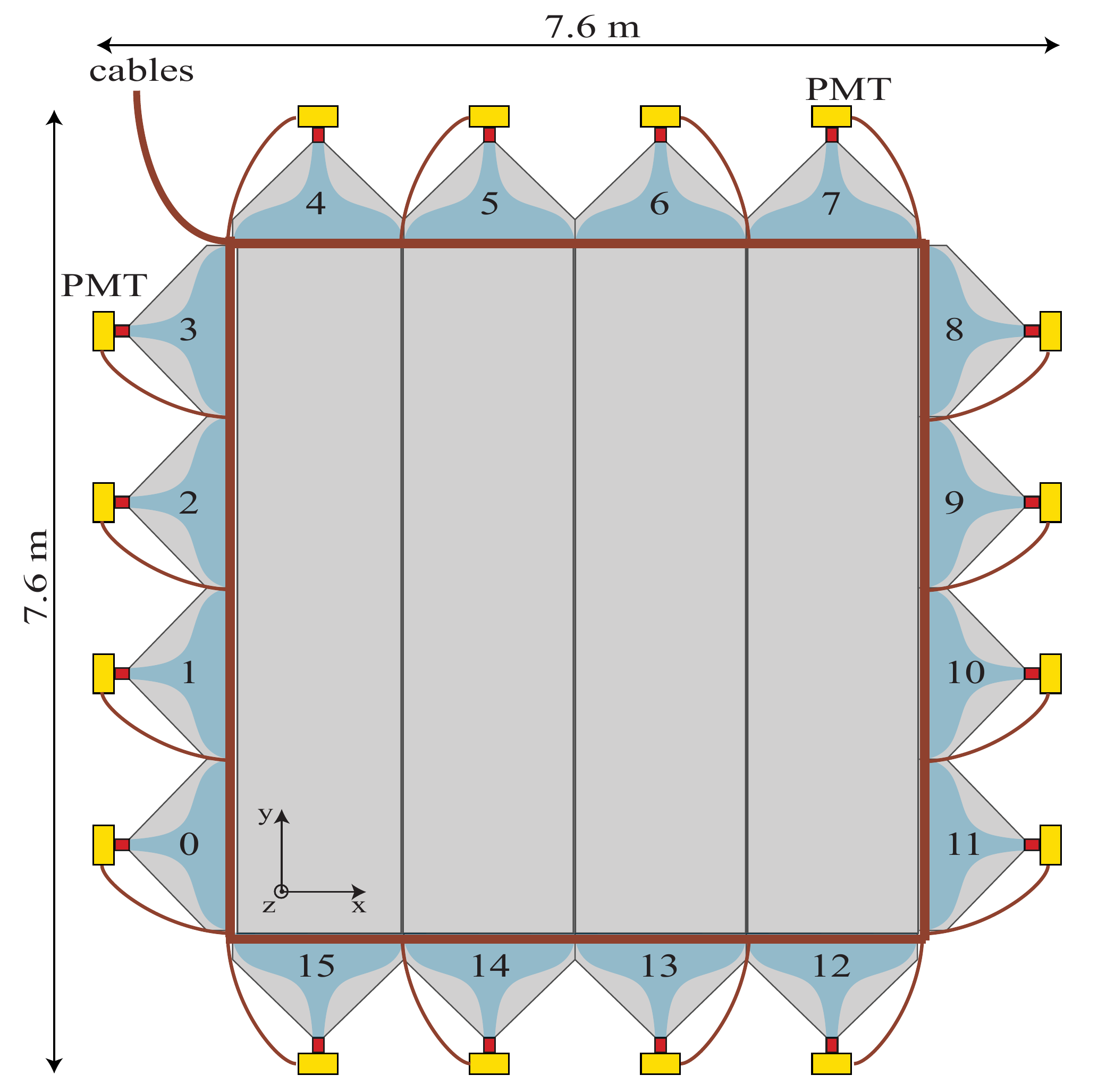}
\caption{Schematic view of a plastic scintillator strip wall.
In addition to the 4 modules clearly visible in the drawing, there are 4 additional modules placed below them
in a perpendicular orientation, going for example between the PMTs numbered 0--11.
} \label{wall_schematic}
\end{figure}

A basic unit (module) of the TT consists of 64 strips
read out by WLS fibres routed at both ends to two 8$\times$8-channel photodetectors placed inside the end-caps
(Fig.~\ref{endcap_schematic}).
Four modules are assembled together to construct a tracker plane
covering $6.7\times 6.7$~m$^2$ sensitive surface.
Two planes of 4 modules each are placed perpendicularly to form a tracker wall providing 2D track information (Fig.~\ref{wall_schematic}).
The inherited OPERA Target Tracker contains 62 walls, with a total number of
31744 scintillating strips for 63488 electronic channels.

\begin{figure*}[hbt]
\centering
\includegraphics[width=0.9\linewidth]{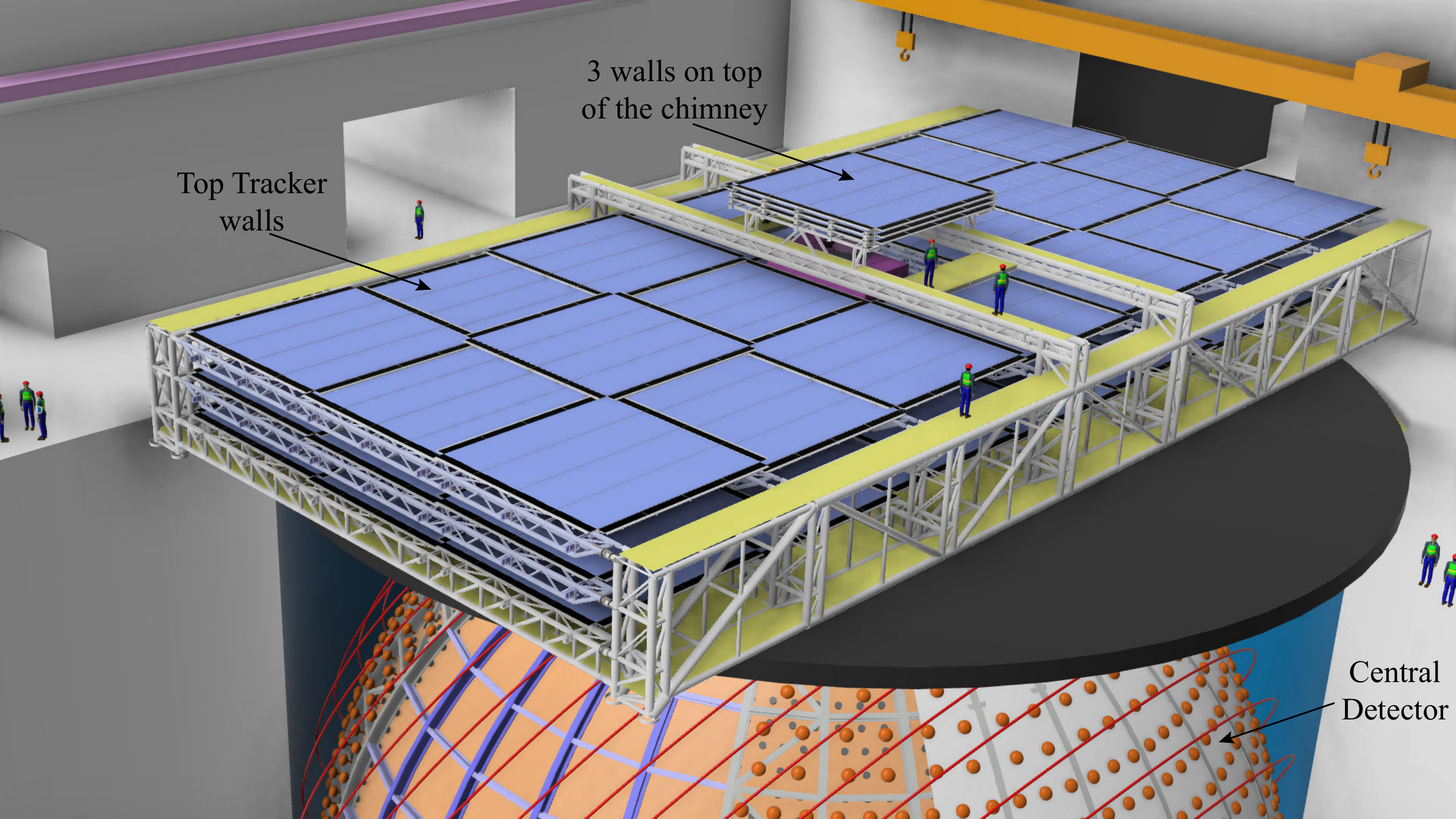}
\caption{Schematic view of the JUNO Top Tracker on top of the Central Detector.}
\label{ttjuno}
\end{figure*}

The JUNO Top Tracker walls, contrarily to their utilisation in OPERA, have to be placed horizontally as  shown by Fig.~\ref{ttjuno}.
A bridge is placed on top of the JUNO Central Detector serving as support for the TT structure.
The three layers of the TT are placed on this bridge, with overlapping walls to limit the dead zones.
The distance of 1.5~m between each layer is a compromise between the muon track reconstruction accuracy and the total size of the needed supporting mechanical structure.
The position of the bottom layer is defined by the minimum distance between the surface of the water of the Cherenkov veto detector and the TT.
Three walls over a distance of 21~m are placed in one horizontal direction and 7 walls over a distance of 40~m in the other direction, leaving a hole in the middle for the JUNO chimney (20 walls per layer).
In this way, 60 walls are used, while the modules of the remaining two walls are reshuffled and placed on top of the chimney, part of the JUNO detector not surrounded by the Water Cherenkov veto detector.
To have three full layers just on top of the chimney, one full TT wall is placed in the lowest part and two partial TT walls are placed in the middle and upper parts, each of those separated by 23~cm.
The partial walls above the chimney contain only the 2 central modules in each layer, for a total of 4 modules per
wall, and thus fully covering only the centre of the chimney.

According to their position, the modules of each wall are placed facing the top or bottom directions in a way to have their end caps, and, in particular, their electronics, accessible from outside.

\section{Trigger rate} \label{radioactivity}

While, as discussed previously, the goal of the TT is to detect atmospheric muons, these are not
the most abundant particles reaching the TT.
Indeed, the natural radioactivity of the elements composing the TT (${0.29 \pm 0.05}$~Hz/strip, or
$\sim 9$~kHz in the entire detector)~\cite{Adam:2007ex}
is larger than the expected $\sim 4$~Hz of muons crossing
the entire Top Tracker.
This natural radioactivity rate is, however, also about three orders of magnitude smaller than the events induced in the TT due
to the radioactivity of the rock of the cavern where the detector will be installed.
The trigger rate of the TT will be directly tied to this rock radioactivity, which will be
described in this section.
To cope with it, specific electronics described in
Sec.~\ref{electronics} will be needed.
Due to the sizeable difference in rate between the detector natural radioactivity and that from the cavern,
along with an expected similarity between the two types of signature,
the detector's natural radioactivity will be neglected for the rest of this paper.

During the prospecting of the JUNO site, rock samples were collected to determine the
content of radioisotopes.
It was determined\footnote{Measurements realised by Ph. Hubert (CENBG Bordeaux) using a Ge detector.}
that the rock radioactivity at the JUNO site is about two orders of
magnitude higher than that in the LNGS where the detector had been previously used, as shown
in Tab.~\ref{radio}.
In addition to the higher radioactivity from the cavern, in JUNO the TT will also be directly exposed to the rock, while the OPERA TT modules were shielded by lead/emulsion bricks.

\begin{table}[hbt]
\centering
\caption{Rock radioactivity in LNGS and in the JUNO site.}
\begin{tabular}{c|c|c}
  \multirow{2}{*}{Isotope}              & \multicolumn{2}{|c}{Activity (Bq/kg)}   \\
                                               & LNGS & JUNO site  \\ \hline
  $^{40}$K                              &  $26 \pm 2$                                  &   $1340 \pm 50$     \\
$^{238}$U                              &  $1.8 \pm 1$                                 &   $110 \pm 10$    \\
$^{232}$Th                              &  $1.5 \pm 1$                                &   $105 \pm 10$   \\
\end{tabular}
\label{radio}
\end{table}

\begin{figure}[hbt]
    \centering
    \includegraphics[width=\linewidth]{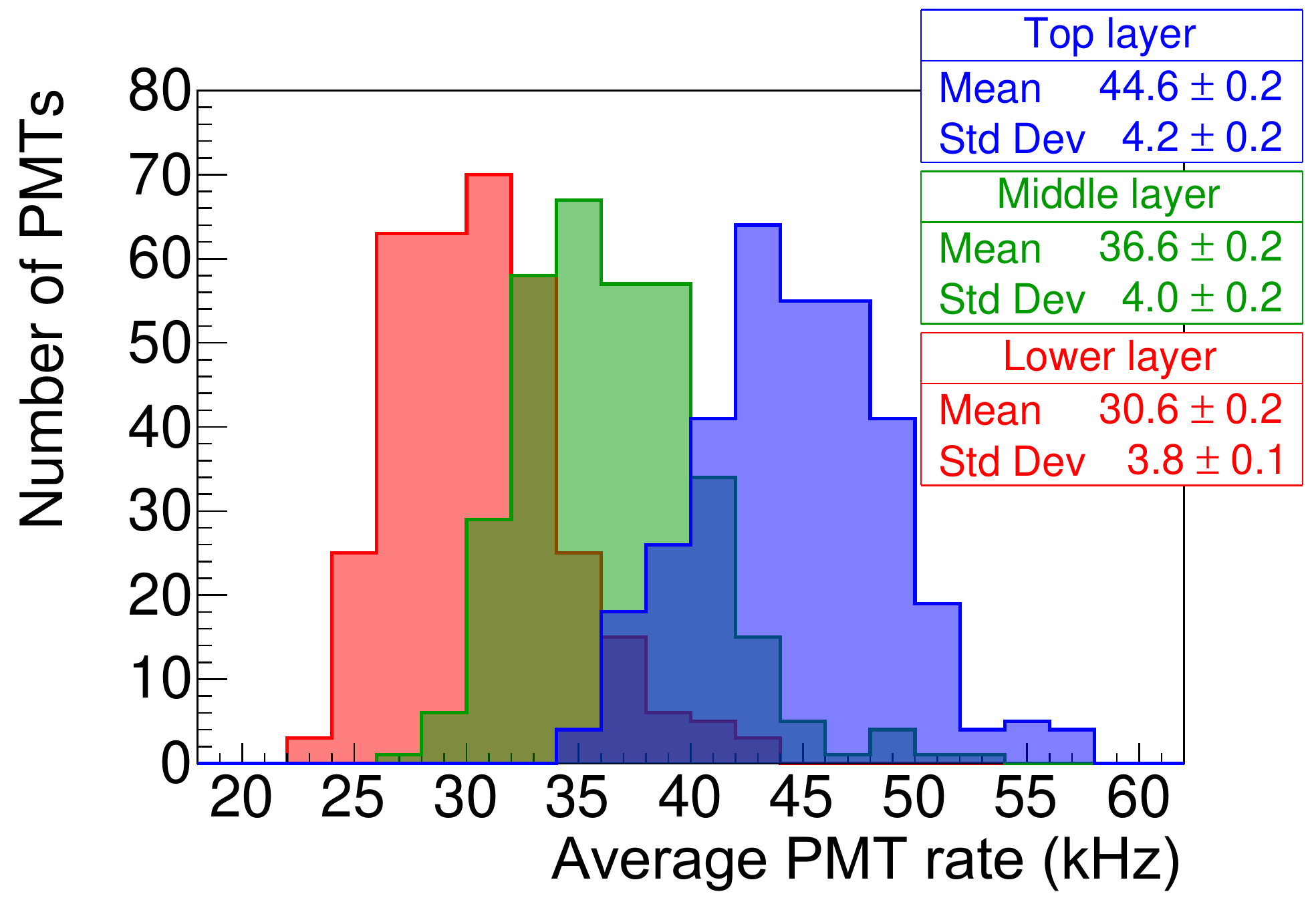}
    \caption{Trigger rate per \mbox{MA-PMT} for the three TT layers.
    The vertical axis represents how many \mbox{MA-PMT} have each expected average rate shown in the
    horizontal axis.
    A \mbox{MA-PMT} trigger threshold of $1/3$~p.e.\@{} is assumed.}
    \label{rate}
\end{figure}

The \mbox{MA-PMT} trigger rate induced by the rock radioactivity is estimated using the
JUNO official simulation software~\cite{Li:2018fny}, where the TT modules are placed as previously described.
In this study decays were individually simulated from the surrounding rock (within a 50~cm thickness from the rock surface) and
the decay particles are then propagated using GEANT4~\cite{Agostinelli:2002hh}.
All energy deposited in the plastic scintillator is then converted into the expected signal at the \mbox{MA-PMT}
based on the detector description in~\cite{Adam:2007ex}.
The \mbox{MA-PMT} is triggered when the collected signal is higher than the threshold of $1/3$~photoelectron (p.e.).
Based on these simulations,
the average trigger rate per \mbox{MA-PMT} varies between 20~kHz and 60~kHz, as
shown in Fig.~\ref{rate}.
The trigger rate per channel has an average value of about 670~Hz, with some
channels exceeding 1~kHz, which has to be compared with about a 20~Hz rate for the TT operation
in the LNGS.
As is also shown in Fig.~\ref{rate},
the rate observed in each \mbox{MA-PMT} will depend on its position given that the modules closer
to the rock will be more exposed to radiation than the more central ones, and the upper layers provide
some shielding for the lower ones.

\begin{figure}[hbt]
  \centering
  \includegraphics[width=\linewidth]{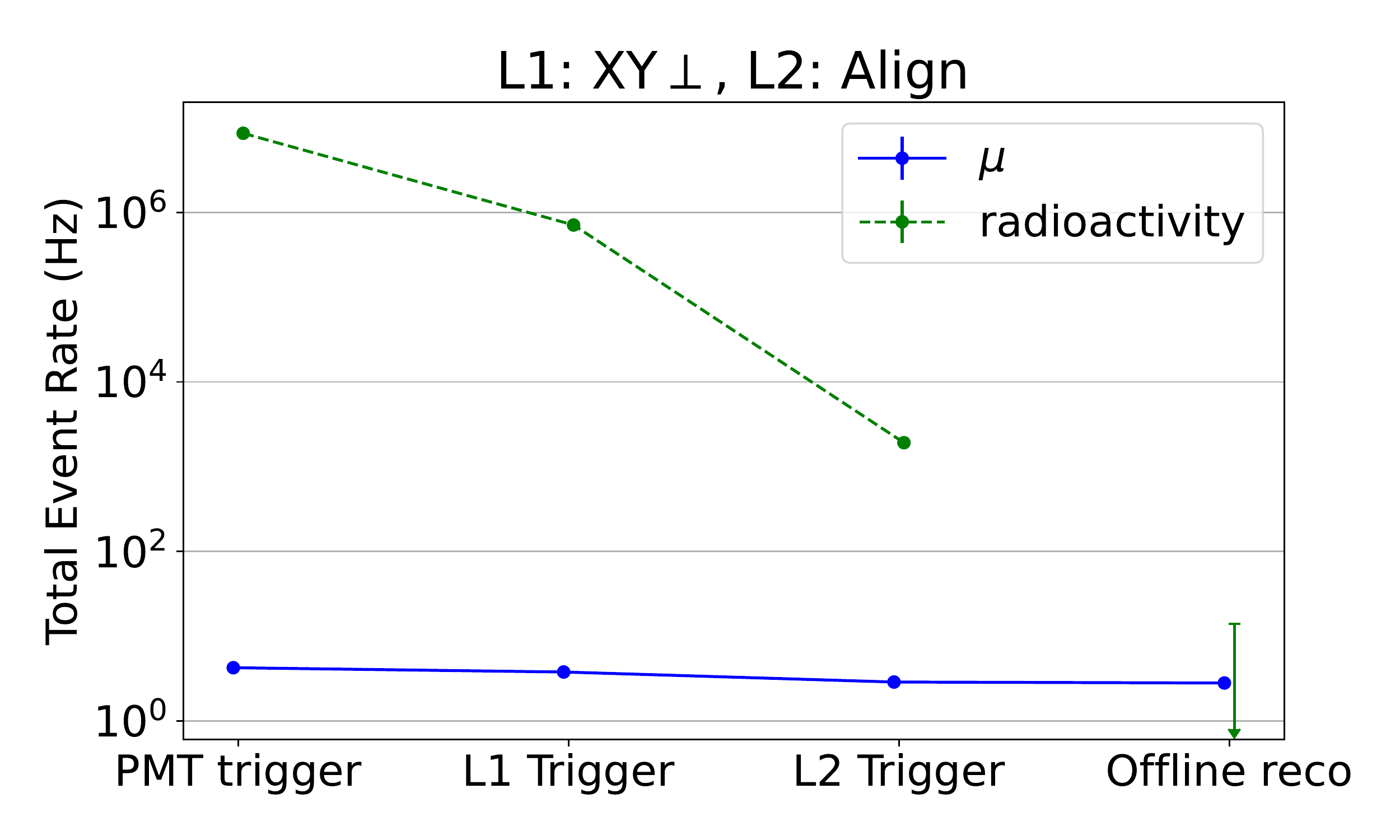}
  \caption{Total event rate of the full detector at various trigger levels, from the \mbox{MA-PMT} trigger rate up
  to the rate after events are reconstructed by the Top Tracker offline software,
  due to radioactivity in the surrounding rocks (green) and due to atmospheric muons (blue).
  The L1 selection criteria requires an X and Y coincidence involving 3 out of 4 \mbox{MA-PMTs} at the end of
  the corresponding modules in the perpendicular X and Y directions.
  The L2 selection criteria requires the L1 coincidences to be on 3 walls of the Top Tracker located
  on different layers.
  }
\label{fig:L2_rate}
\end{figure}

The high trigger rate at the \mbox{MA-PMT} level expected due to the rock radioactivity
imposes strong constraints on the
new TT acquisition system.
Given that the radioactivity produced by the decay of $^{40}$K, $^{232}$Th, $^{238}$U and their
decay products will not typically traverse more than one layer of scintillator,
it is possible to reduce the amount of data generated during the JUNO experiment by requiring
a time coincidence between triggers related to \mbox{MA-PMTs} along perpendicular modules in the same wall.
Additionally, to further reduce the produced data, it is also required
that these time coincidences are found in the three aligned TT walls of different layers.
With these coincidence criteria the full detector event rate can be reduced from
about 8~MHz, which corresponds to the total rate over the full detector from the per PMT rates shown
in Fig.~\ref{rate},
to about 2~kHz, as shown in Fig.~\ref{fig:L2_rate}, which can be handled by the acquisition system.
The two levels of time coincidence described correspond to the L1 and L2 triggers.
The L1 trigger relates
to the time coincidences within a single wall while the L2 trigger does the same at the level of the entire TT.
The implementation of these coincidence criteria in the electronics is discussed in the following sections.

It is worth noting that in Fig.~\ref{fig:L2_rate} were chosen the L1 and L2 algorithms
that are the most likely to be used in the Top Tracker, however the final decision has not yet been made.
The L1 trigger, which is discussed in Sec.~\ref{cbelectronics}, provides about an order of magnitude reduction
of the rate by requiring 100~ns coincident \mbox{MA-PMT} triggers in the two perpendicular planes of any given wall,
and with at least one of the planes having the \mbox{MA-PMTs} on both sides of the module triggered.
The L2 trigger, which is discussed in Sec.~\ref{gtbelectronics}, provides another reduction factor of
about 400 to the event rate by requiring at least 3 aligned walls on different layers of the detector
to have produced a L1 trigger within a 300~ns sliding time window.
In case an alternative algorithm is used in the L1 and L2 electronics cards, due to
difficulty in implementing them, and the
data reduction is not as efficient as discussed here, it is foreseen that the
data acquisition system will be able to apply online additional criteria
to reduce it to the rate described above.

It is also useful to note, as shown in Fig.~\ref{fig:L2_rate} that the radioactivity rate can be further
reduced offline by performing a full 3D reconstruction of the muon track,
rather than just requiring walls to be aligned.
This reconstruction is presented in Sec.~\ref{sec:reconstruction} and corresponds to the
``offline reco'' level on the aforementioned figure, for which only a 68\% upper limit is shown for radioactivity.
Any potential remaining radioactivity induced events after the reconstruction
can still be further reduced by requiring a correlated signal between the Top Tracker and
the central detector or the water Cherenkov veto detector, after that the
impact of the radioactivity becomes negligible
compared to the atmospheric muon rate.

\section{Electronics} \label{electronics}

\begin{figure}[hbt]
\centering
\resizebox{.95\linewidth}{!}{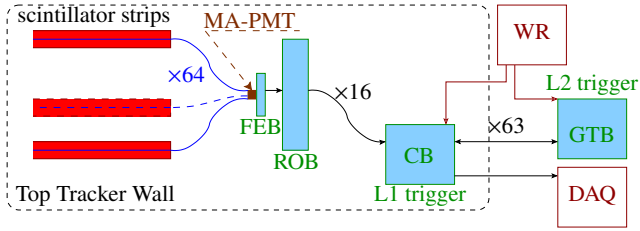}
  \caption{TT electronics chain.
  All electronics cards developed for the TT are shown in blue and are the
  front end board (FEB), readout board (ROB), concentrator board (CB) and global trigger board (GTB).
  The FEB and ROB are installed in each module end-cap.
  The FEB is connected to the \mbox{MA-PMT} that
  observes the light collected by WLS fibres in the 64 scintillator strips of each module and to the ROB.
  The ROB is connected to the FEB and to the CB of that specific wall.
  The CB is located at the centre of the wall and is connected to all 16 ROBs of the wall.
  The CB is also connected to the white rabbit (WR) switch, to the data acquisition system (DAQ) and to the GTB.
  The GTB, in addition to being connected to the 63 CBs of the TT, will also be connected to the WR switch.
  }
\label{allelectronics_zoom}
\end{figure}

\begin{figure}[hbt]
\centering
\includegraphics[width=\linewidth]{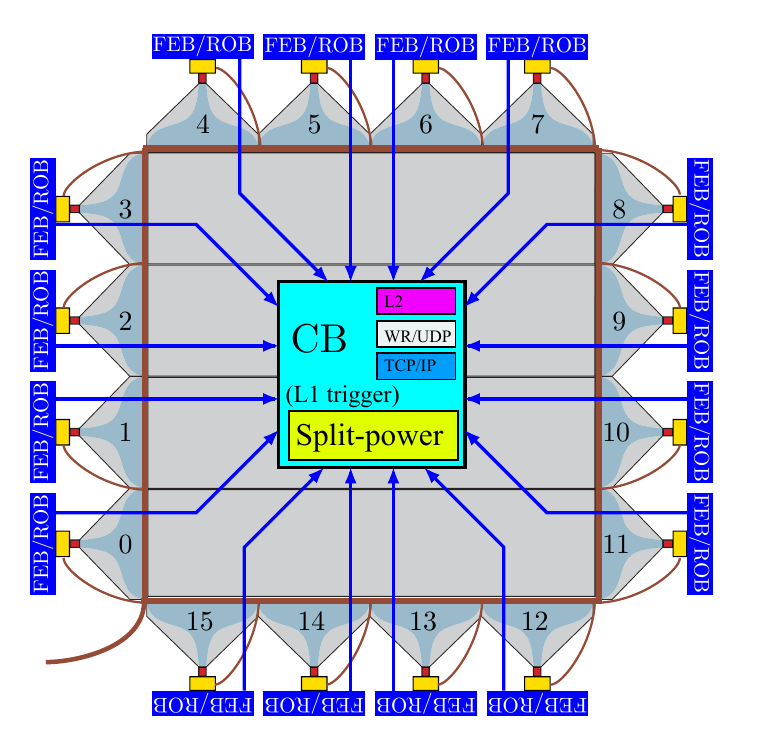}
\caption{Schematic view of the TT wall electronics.
  The FEB and ROB at each end-cap are connected to the CB which performs the L1 trigger.
  The CB connections are shown as boxes in the top right corner of the CB box.
  In particular the CB is connected to the L2, to the white rabbit (WR) switch via UDP and to the data acquisition system via TCP.
  Next to the CB in the wall is also located the split-power board that distributes the power for all ROBs and the
  CB in each wall.
  }
\label{allelectronics}
\end{figure}

The global electronics conceptual readout chain is shown in Fig.~\ref{allelectronics_zoom}.
As it was previously mentioned, the TT is a set of 63 identical walls with similar mechanical constraints per wall which leads to the development of individual electronics per wall. 
This electronics chain is implemented at the TT wall level as shown in Fig.~\ref{allelectronics}.
Due to the limited thickness of the module end-caps (3~cm) the PMT readout had to be split in two boards: the front end board (FEB) directly connected to the \mbox{MA-PMTs} and the readout board (ROB) connected to the FEB and the wall concentrator board (CB).
Both FEB and ROB are placed in the module end-caps next to the \mbox{MA-PMT}.
The CB, located in the centre of each wall, configures and controls the 16 FEBs and ROBs in a wall,
and performs the data readout for the full wall which will subsequently be transferred to the data acquisition
system by the CB.
The \mbox{MA-PMTs} are read out by the 64-channel MAROC3~\cite{Blin:2010tsa} ASIC placed on the FEB.
The ROB is placed after the FEB to receive the \mbox{MA-PMT} data and to send them to the CB that gathers the data from the 16 sensors of a TT wall.
A split-power placed at the level of the CB powers the \mbox{MA-PMTs}, the FEB, the ROB and the CB.

The charge and hit information are given by the MAROC3 chip together with an ``OR'' trigger signal used by the CB for time-stamping.
The CB identifies \mbox{x~--~y} correlations at the level of the wall (L1 trigger) 
with a sliding time window of 100~ns.
In the absence of correlations in a wall the CB will send back a RESET signal to the ROB in order to reduce the
dead time of the FEB MAROC3 (between 7--14~\textmu{}s).
In case a coincidence is detected the CB sends this information to a Global Trigger Board (GTB), which further searches for correlations at the level of all three TT layers (L2 trigger) with a sliding window of 300~ns.
The GTB sends back a validation or a rejection signal to the CBs that, correspondingly,
either sends the related data to the JUNO data acquisition system or
issues a reset signal to the relevant ROBs in about 1~\textmu{}s.
Both trigger levels are essential to significantly reduce the TT data rate, as discussed previously, and to reduce
the detector dead time of the FEB MAROC3.

\subsection{Front End Board (FEB)} \label{febelectronics}

\begin{figure}[htb]
\centering
\includegraphics[width=\linewidth]{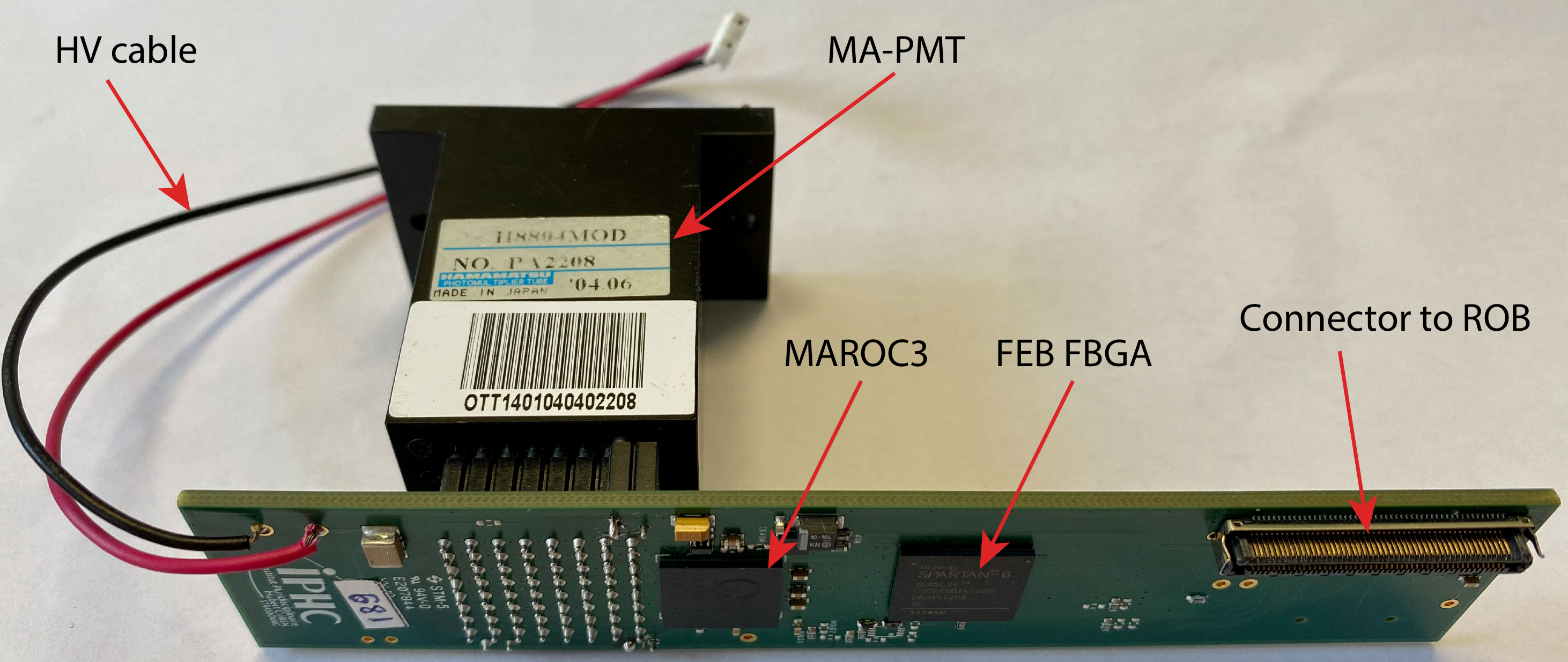}
\caption{The FEB connected to a \mbox{MA-PMT}.}
\label{front_end_PCB}
\end{figure}

The FEB is an 8-layer PCB, 3~cm~$\times$~13~cm in size, carrying the MAROC3 and a small Spartan6 FPGA used to serialise the hit signals from 64 parallel outputs to 8 serial links.
This operation allows the ROB to obtain the information of which channels have triggered, which is then used for online zero suppression if that is the set acquisition mode.
The board is directly plugged to the \mbox{MA-PMT}, as shown in Fig.~\ref{front_end_PCB}.
The lines from the \mbox{MA-PMT} to the MAROC3 inputs are well separated and protected from external noise sources by four ground planes in the PCB.

The readout electronics of the Top Tracker is based on a 64-channel ASIC, the MAROC3 designed by
the Omega laboratory~\cite{Blin:2010tsa}.
The chip is provided in a 12~mm~$\times$~12~mm TFBGA package with 353~pins.
The main characteristics of this chip are:
\begin{itemize}
  \item Configurable gain compensation (called gain equalisation factor in the following, g.e.f.), varying from 0 to 3.9, to equalise the anode-to-anode gain differences of the \mbox{MA-PMT} channels~\cite{Adam:2007ex}.
    The MAROC3 is equipped with an adjustable gain system incorporated in the preamplifier stage delivering an identical amplification factor to both, the fast shaper (trigger arm) and slow shaper (charge measurement arm), of each channel.
  \item Delivery of a global low noise auto-trigger with 100\% trigger efficiency for minimum ionising particles,
    using a threshold as low as 1/3~p.e., corresponding to 53~fC for a total gain of $10^6$.
  \item Delivery of a charge proportional to the signal detected by each \mbox{MA-PMT} channel in a dynamic range up to 5~pC ($\sim 30$~p.e.\@{}\footnote{A gain of $10^6$ is assumed in all charge to p.e.\@{} conversions in this paper.}).
\end{itemize}

\begin{figure*}[hbt]
\centering
\includegraphics[width=.7\linewidth]{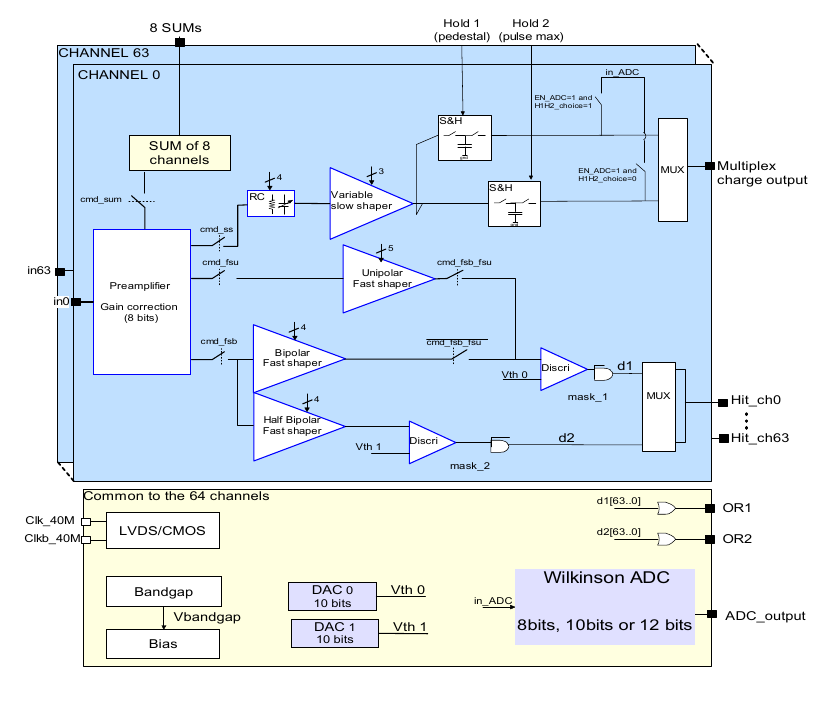}
\caption{Architecture of MAROC3 ASIC~\cite{Blin:2010tsa}.}
\label{maroc3}
\end{figure*}

As mentioned in the list above, each of the 64 channels comprises a low noise variable gain preamplifier that feeds both a trigger and a charge measurement arm (Fig.~\ref{maroc3}).
The auto-trigger includes a fast shaper followed by a comparator.
This signal is compared with an adjustable threshold to deliver a trigger signal for each channel.
The trigger decision is provided by the logical ``OR'' of all 64 channels.
A malfunctioning channel can be disabled from the trigger decision through a mask register.

The charge measurement arm consists of a slow shaper followed by a Track\&Hold buffer.
The slow shaper has a gain of 121~mV/pC (equivalent to 19~mV/p.e.\@{}) and a long peaking time to minimise the sensitivity to the signal arrival time.
Upon a trigger decision, charges are stored in capacitors and the 64 channels are readout sequentially through a single analogue output, called OutQ, at a 10~MHz frequency.
The charge output can also be directly read with an 8, 10 or 12~bits internal Wilkinson ADC.
For the TT the 8~bit readout was selected to be used with the Wilkinson ADC to reduce the readout deadtime.
The output clocking frequency is 80~MHz.
In the 8~bit case all 64~channels are read in less than 14~\textmu{}s,
and while the channels are being read the MAROC3 will not be available to measure another charge resulting
in detector deadtime.
This time can be reduced if a reset signal is received from the CB via the ROB, at which point the readout and
transfer process is stopped and the MAROC3 is ready to start again charge measurement.
It's worth noting that this deadtime does not affect the trigger system discussed in the previous paragraph.

\begin{sloppypar} 
The MAROC3 chip consumes approximately 220~mW (3.5~mW/channel), with a small variation depending upon the gain correction settings and ranges.
\end{sloppypar}

The variable gain system is activated by eight switches which allow setting the g.e.f.\@{} varying from 0 to 3.9.
By turning off all current switches of a channel, the g.e.f.\@{} is zeroed and the channel is disabled.
For a g.e.f.\@{} of 1, the switches are set to 64.

The trigger efficiency can be measured as a function of the injected charge for each individual channel.
100\% trigger efficiency is obtained for input charge as low as 0.1~p.e., independently of the g.e.f.
With the trigger threshold set externally and common to all channels, the output spread among the 64 channels can be carefully controlled.
It is found to be around 0.01~p.e., an order of magnitude smaller than the useful threshold level.

Upon a trigger decision, all the capacitors storing the charge are read sequentially through a shift register made by D-flip-flop.
The pedestal level is less than 1/3~p.e.\@{} considered as the final  TT trigger level.

\begin{figure}[hbt]
\centering
\includegraphics[width=\linewidth]{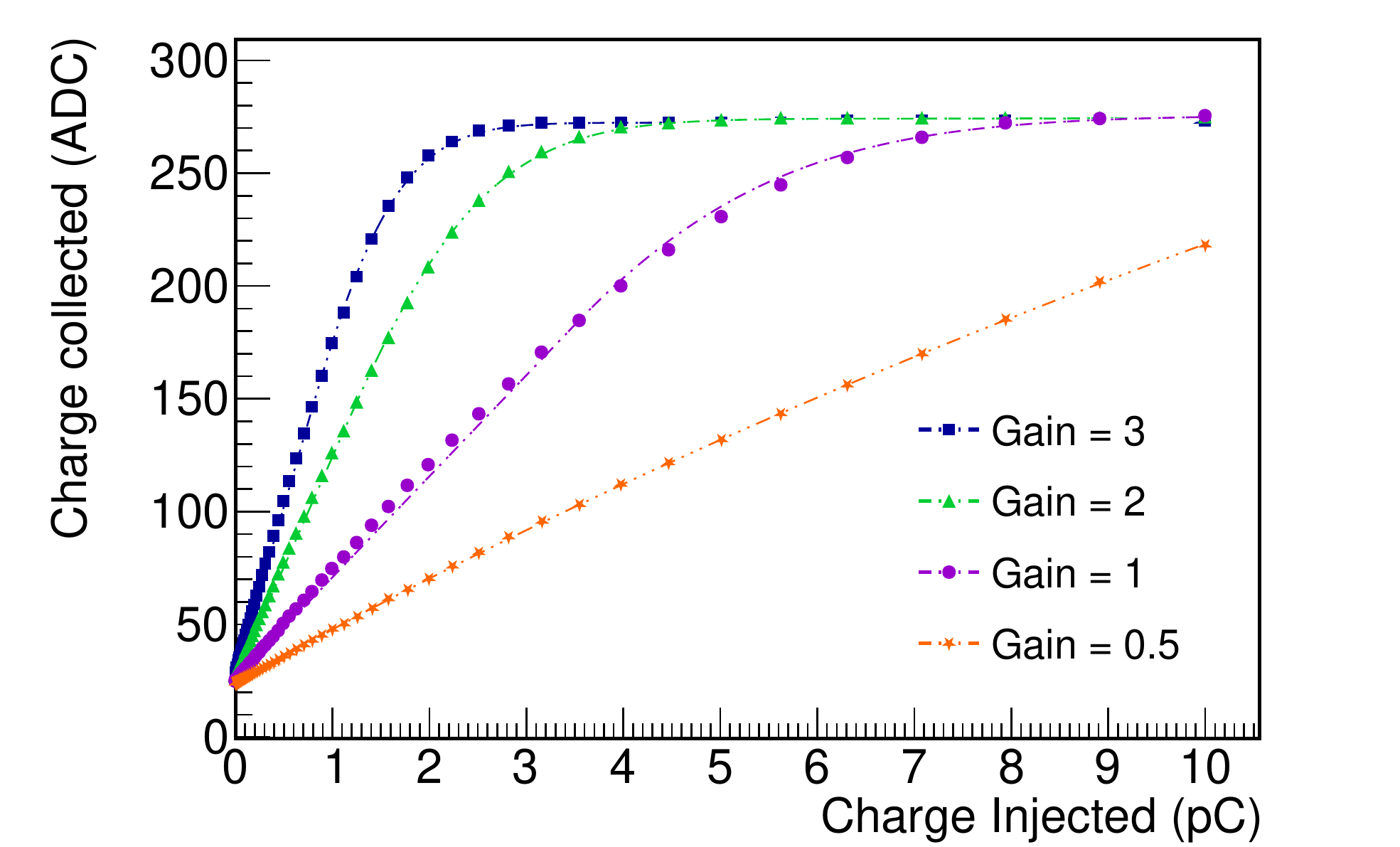}
\caption{Linearity of the charge measurement as function of the input charge for a g.e.f.\@{} set to 0.5, 1, 2 and 3.}
\label{vers3_charge_linearity_vs_gain}
\end{figure}

The reproducibility  in the charge measurement has been determined for all channels of the same chip and found to be better than 2\% over the full range 1--10~pC for a g.e.f.\@{} of 1, corresponding to 1--35~p.e.\@{}
(Fig.~\ref{vers3_charge_linearity_vs_gain}).

Cross-talk related to the ASIC has been carefully considered.
Two main sources of cross-talk have been identified.
The first one comes from a coupling between the trigger and the charge measurement arms and has been determined to be lower than 0.1\%.
The second one affects the nearest neighbours of a hit channel, where a cross-talk of the order of 1\% has been measured.
This effect is mainly due to the routing of traces in the board and is negligible for far channels.

The FEB contains buffer amplifiers for the differential charge output signals of the MAROC3 and logic level translators for the digital signals.
The FEB is connected to the ROB board by means of one 100 pins coaxial connector.
This two-board arrangement conforms to the mechanical constraints of the TT module’s end-cap geometry.

\subsection{ReadOut Board (ROB)} \label{robelectronics}

\begin{figure*}[hbt]
\centering
\includegraphics[width=0.7\linewidth]{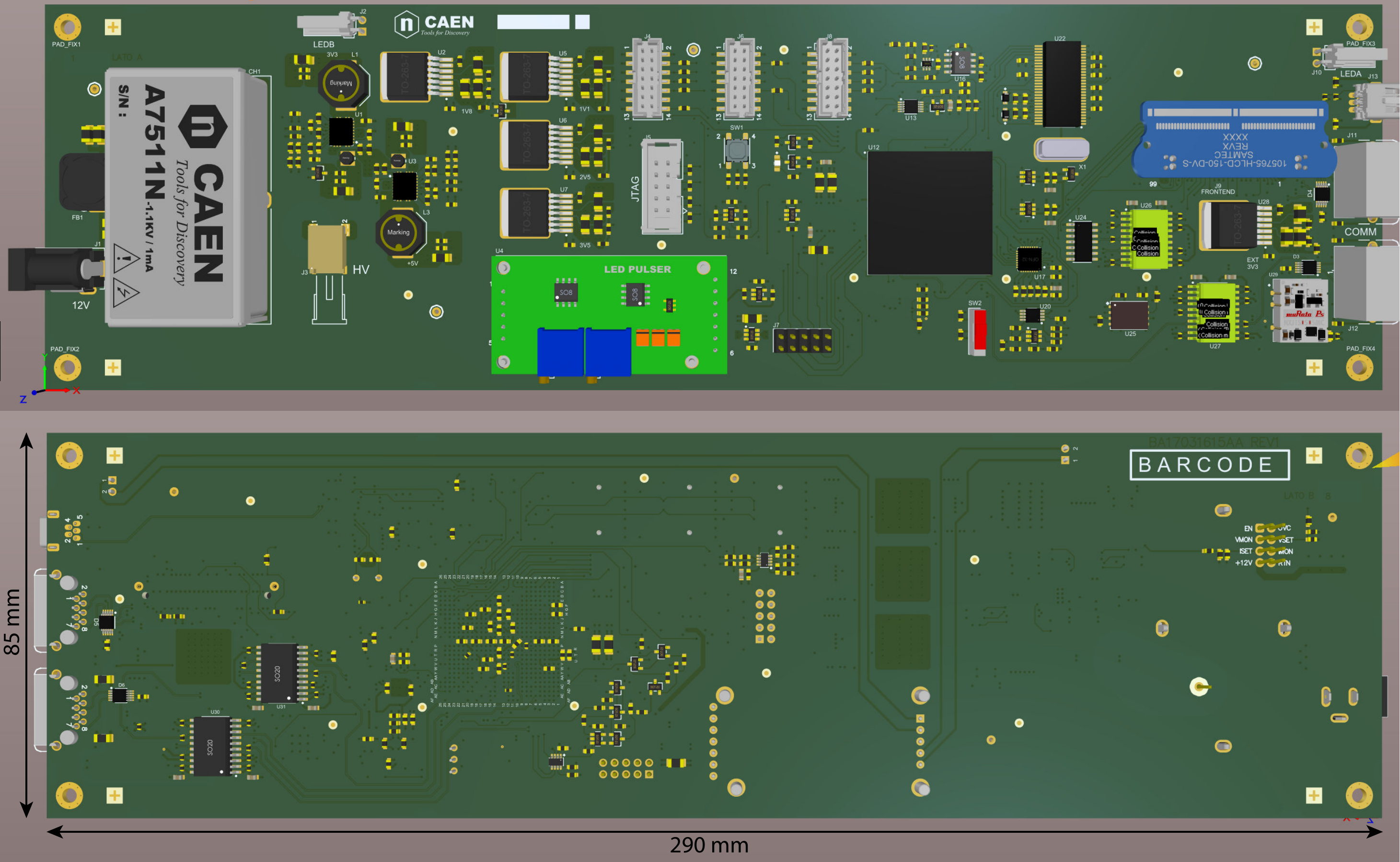}
\caption{Top and bottom sides of the ROB.}
\label{rob}
\end{figure*}

The ROB manages the FEB power, acquisition and slow control.
It is an 8-layer PCB developed in collaboration with CAEN (mod.\@{} A1703) based on an ALTERA Cyclone~V~GX FPGA.
A picture of the PCB top and bottom views is shown in Fig.~\ref{rob}.

In standard acquisition mode, the FPGA state machine waits for the detection of a FEB ``OR'' signal. 
Following the detection of this signal, a Hold signal is issued, for charge measurement in the MAROC3 chip, with a tunable delay and a time jitter lower than 1~ns.
In this way, a charge measurement precision better than 1\% is achieved.
The triggered channel pattern
is also acquired and used for zero-suppression, as discussed previously.
The ROB features a 12~bit Flash ADC converter \mbox{AD9629BCPZ-65}~\cite{link:FADC} by Analog Devices for the read out of the OutQ signal from MAROC3 chip (analogue output proportional to the charge to be measured).
In such a way, the readout of one event is performed in about 7~\textmu{}s.
The readout of the charge converted by the internal Wilkinson ADC discussed previously is also possible,
however that will require up to the previously mentioned 14~\textmu{}s.
Due to this difference in the time required for the charge readout,
the Flash ADC is expected to be used for charge measurement during data taking.

The ROB also hosts a High Voltage module to supply the \mbox{MA-PMT}, which is operated in a range from $-1000$ to $-800$~V, with a current lower than 500~\textmu{}A, and a test pulse unit for calibration purposes.

\begin{sloppypar} 
The HV module is the \mbox{A7511N} DC--DC converter by CAEN~\cite{link:A7511} with a ripple lower than 10~mV.
Tests carried out on a detector prototype have shown that the DC--DC converter noise contribution to the \mbox{MA-PMT} output signal is lower than 1~mV.
\end{sloppypar}

The test pulse generators, illuminating the fibres in the end-caps of the TT modules, are the same used in OPERA~\cite{Adam:2007ex}.
The generator produces short pulses, 20~ns wide with adjustable amplitude, powering blue LEDs which excite the WLS fibres exiting the scintillator strips.
In the calibration procedure the pulse amplitude is set in such a way to simulate the single photoelectron signal on the \mbox{MA-PMT} while the HOLD signal is issued with an adjustable delay with respect to the LED driver pulses, emulating the acquisition of standard signals.   
Pedestals can be measured with random emission of HOLD signals inside the ROB.
These various data acquisition modes will be further described in Sec.~\ref{operation}.

At nominal working conditions on a detector prototype, an operating current of 350~mA has been measured at the nominal operating voltage of 12~V supplied to the ROB, powering also the FEB and the \mbox{MA-PMT}.

Finally, the ROB includes the possibility to remotely re-flash the FEB Spartan6 FPGA, to facilitate remote system operation.

\subsection{Concentrator Board (CB)} \label{cbelectronics}

At the centre of each TT wall is the CB,
as shown in Fig.~\ref{allelectronics}.
Its main role is to aggregate the 16 incoming ROB data streams, perform data selection (Level~1 trigger), and forward the reduced data stream to the data acquisition system.
All 63~CBs present in the detector operate in parallel and fully independently, but with a well-synchronised timing distribution system based on the White-Rabbit (WR) standard~\cite{6070148}.

\subsubsection{Interfacing to 16 ROBs} \label{cbmux}

Placing the CB at each wall makes it possible to reduce the number of connections to the data acquisition system from 992 (counting one for each ROB) to just 63 for the whole detector.
Such wall-centric organisation is shown in Fig.~\ref{allelectronics}.

The board provides four remote connections: one copper cable for the system power supply at $+12$~V (PWR) and three optical fibres for long-distance data links. One of the optical links implemented in the CB is used for the connection to the Detector Control System using the User Datagram Protocol (UDP) and for the WR time synchronisation system.
Another optical link implements a dedicated ethernet connection (TCP) to the data acquisition system  and the final link connects the CB to the GTB (see Sec.~\ref{gtbelectronics}) for the Level~2 trigger selection.

All local connections between ROBs and the CB are implemented using Cat5e cables.
Associated with the CB board is a small power-distribution card referred to as Split-Power Board that is briefly described in Sec.~\ref{splitpower}.

\subsubsection{L1 trigger selection} \label{cbL1}

As explained in Sec.~\ref{radioactivity}, the data stream produced by a ROB card is dominated by radioactivity induced background events.
At the expected 50~kHz average ROB trigger rate, the resulting data rate from a thousand ROBs would be unmanageable at the data acquisition level.

As discussed previously, a reduction of the data rate by an order of magnitude can be achieved using the CB to select only events in \mbox{x~--~y} coincidence (L1 trigger level).
A coincidence is defined here as a set of 2, 3 or 4~FEB/ROB triggers arriving in the same time window of 100~ns and associated with two perpendicular modules in the same wall.
The L1 selection algorithm is implemented as a sliding window on the signals from all 16~ROBs.
To protect from classification ambiguities, any coincidence detected within this window will result in accepting all ROB data inside the window, even if some signals are not correlated to the event.

The triggers arriving at the CB are time-stamped with a resolution of 1~ns using the absolute time reference provided by the WR node implemented in the CB.
For FEB/ROB triggers found to be in coincidence, the CB waits for the associated ROB data packets containing the signal position and its charge value.
After merging the timing and charge information, the modified packets are transferred to the data acquisition system,
provided the L2 trigger did not reject this L1 trigger.
FEB/ROB triggers that do not produce a L1 trigger, or that are rejected by the L2 trigger, are rejected and a reset signal is sent to the corresponding ROB to abort charge digitisation as well as the associated data transfer to the CB.

The overall system synchronisation provided by the WR protocol is necessary for a correct data reconstruction in the data acquisition system and for data analysis.
It is also used for communicating the coincidences to the GTB (a partial timestamp of the L1 trigger is sent from the CB to the GTB) that will confirm or reject them in accordance with the L2 trigger selection applied at the detector level.

\subsubsection{CB implementation} \label{cbhw}

\begin{figure}[hbt]
\centering
\includegraphics[width=\linewidth]{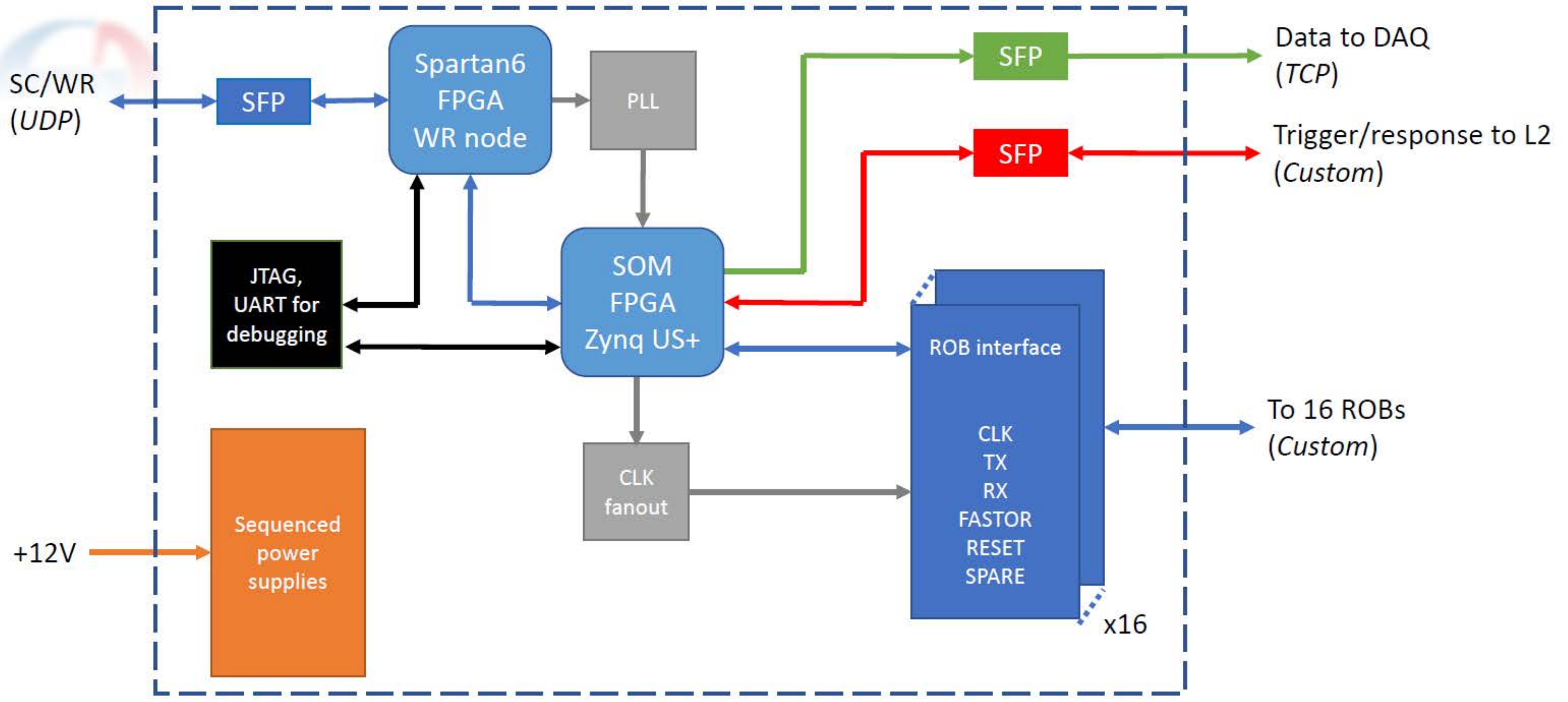}
\caption{CB block diagram.}
\label{cb_block_diagram}
\end{figure}

\begin{figure}[hbt]
\centering
\includegraphics[width=\linewidth]{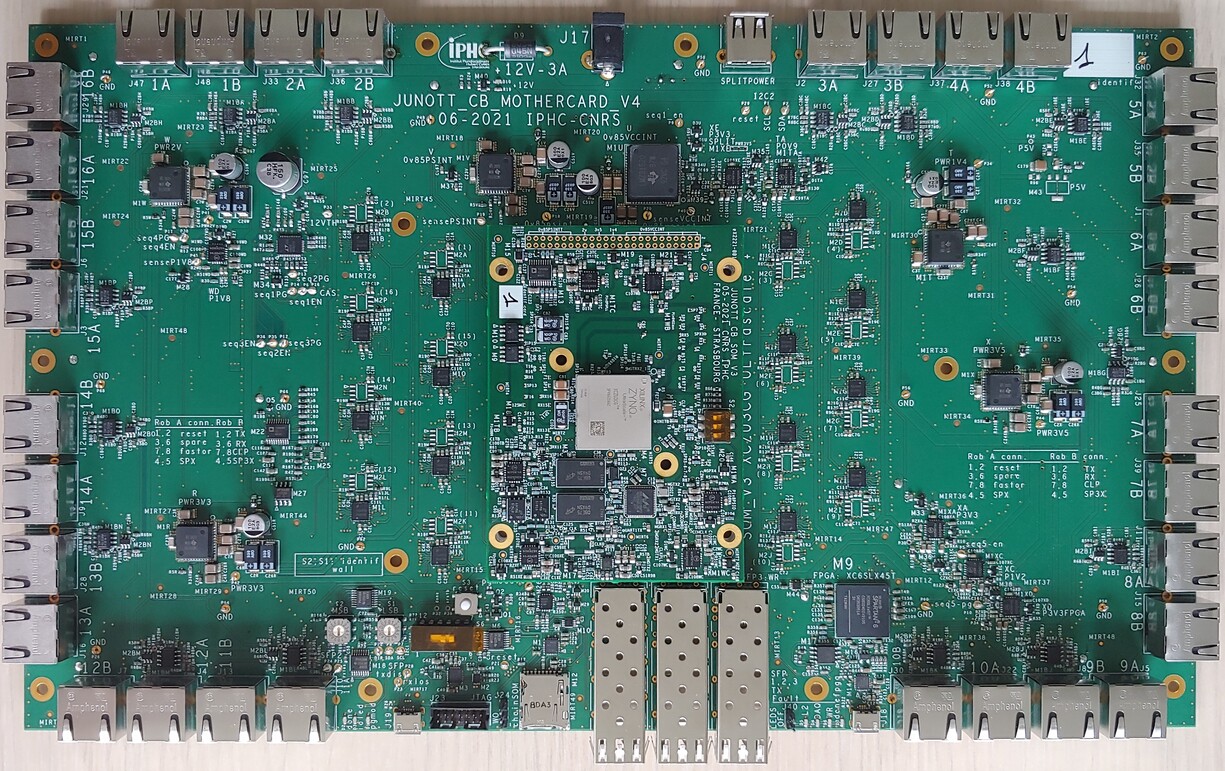}
\caption{CB motherboard and the daughtercard.}
\label{cbv4}
\end{figure}

The CB has been implemented as an FPGA-centred readout system with a motherboard that accommodates a daughter card hosting the processing unit (System-On-Module, SOM).
A block diagram of the CB is shown in Fig.~\ref{cb_block_diagram} and the pre-production prototype is shown in Fig.~\ref{cbv4}.

The CB motherboard is a 34~cm~$\times$~20~cm, 8-layer card that provides all the interface-specific electronics components for supporting the LVDS\footnote{Low Voltage Differential Signalling.}-based connections to the 16~ROBs and the optical connections as described earlier.
It provides a set of all necessary power supply voltages and their sequenced activation to satisfy the requirements of this FPGA-based readout system.
The board also hosts an integrated WR-node that is implemented in a Spartan6 FPGA~\cite{WR_node}.
This FPGA also interfaces the slow-control UDP traffic going to and from the CB with the main FPGA in the daughter card.

\begin{sloppypar} 
The daughter-card is an embedded high-performance, FPGA-based system on module, which controls and processes all incoming and outgoing data streams.
It is a 10~cm~$\times$~7~cm, 16-layer PCB that hosts the Xilinx System-on-Chip (SoC) of the latest generation, Zynq Ultrascale+CG~\cite{Xilinx_Ultrascale}, accompanied by a 2~GB of on-board DDR4 memory.
The programmable-logic part of this FPGA controls the communication with the 16~ROBs and the link to the GTB, both based on custom protocols.
Two of the four available dedicated hardware CPUs in the SoC manage in parallel the TCP traffic going to the data acquisition system and the slow-control.
\end{sloppypar}

\subsubsection{Split-power Board} \label{splitpower}

\begin{figure}[hbt]
\centering
\includegraphics[width=.8\linewidth]{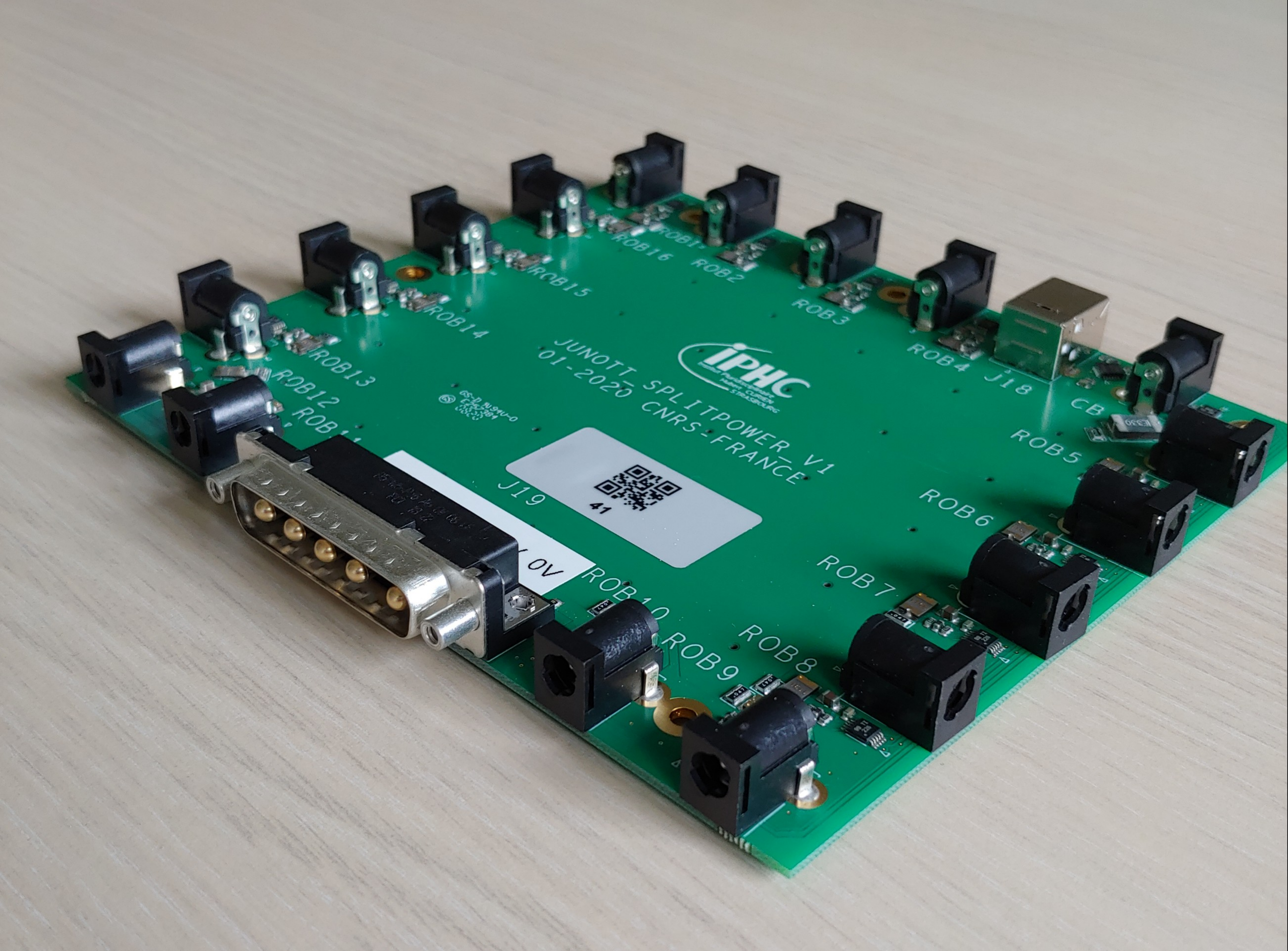}
\caption{Split-power board.}
\label{sp_in_TT_wall}
\end{figure}

The Split-power board, shown in Fig.~\ref{sp_in_TT_wall}, is a dedicated PCB for distributing power supply to all TT wall components.
This board is approximately 15~cm~$\times$~15~cm in size and implemented in 2 copper layers.
The power delivered by a remote power supply is fanned out to 16~ROBs and one CB. The board features re-settable fuses, shunt resistors and 16-bit precision I2C\footnote{Inter-Integrated Circuit.} current/voltage monitors in all the output power branches.
An I2C bus runs between this PCB and the CB to allow the CB to monitor the power consumption of all these readout components.

\subsection{Global Trigger Board (GTB)} \label{gtbelectronics}

The main role of the GTB is to perform a second level (L2) trigger working on the \mbox{x~--~y} coincidences selected by the L1 trigger of each TT wall.
The idea is to accept interesting events depending on the spatial position of the walls involved by the passage of the crossing muons.
The requirement of the algorithm implemented in the GTB is to find an alignment of the \mbox{x~--~y} coincidences over the three TT layers in a given time window.
The second requirement is that this algorithm has to be very fast in order to give an online feedback.
As discussed previously, this level of trigger will be able to further remove the background noise given by radioactivity, correlated electronic noise or \mbox{MA-PMT} dark noise.

The GTB will occupy the top of the trigger chain acting as the final sink of all the trigger requests and the single source of all the trigger validations.
Such topmost position will give the possibility for the GTB to achieve also secondary tasks typical of a trigger root node, such as an injection of a technical trigger (periodically or randomly generated), a throttling of the trigger in case of anomalous situation in which part of the detector is affected by more noise, or a prescale of the trigger for commissioning studies.
GTB will have to take validation/rejection decisions based only on the spatial position of the incoming requests and on the trigger time information associated to them.
Indeed, each trigger request will be associated with a timestamp that specifies the time the corresponding hits have been generated in the FEB. 

Currently, the algorithm for finding aligned \mbox{x~--~y} correlations over the three TT layers has not been decided yet.
Several solutions have been studied in simulation.
The most promising is the coincidence of three aligned walls each on a different layer, however the possibility
of just requiring three walls without any alignment was also considered.
Fig.~\ref{fig:L2_rate} shows the reduction in terms of the total event rate applied by the L2 trigger for the most
promising case.
When only three walls, without any alignment, are required the GTB performs about 20 times worse.
This both underlines the importance of alignment, while still showing that even a significantly less resource
demanding algorithm on the GTB would already provide more than an order of magnitude reduction
to the radioactivity rate passing the L1 trigger.
Further solutions exploiting neural networks on the FPGA are also under study.
Indeed, these tools are widely used for solving the problem of pattern recognition and are implementable in hardware with a limited number of resources with respect to an analytic solution of the problem.
Given the non-homogeneous geometry of the detector, specially with respect to the displaced walls for accommodating the chimney in the middle of the detector, an approximate solution of the alignment problem via pattern matching could lead to much easier implementation still keeping a very good level of efficiency
in comparison to a linear fit of the positions in the centre of all triggered walls.

\begin{figure}[hbt]
\centering
\includegraphics[width=\linewidth]{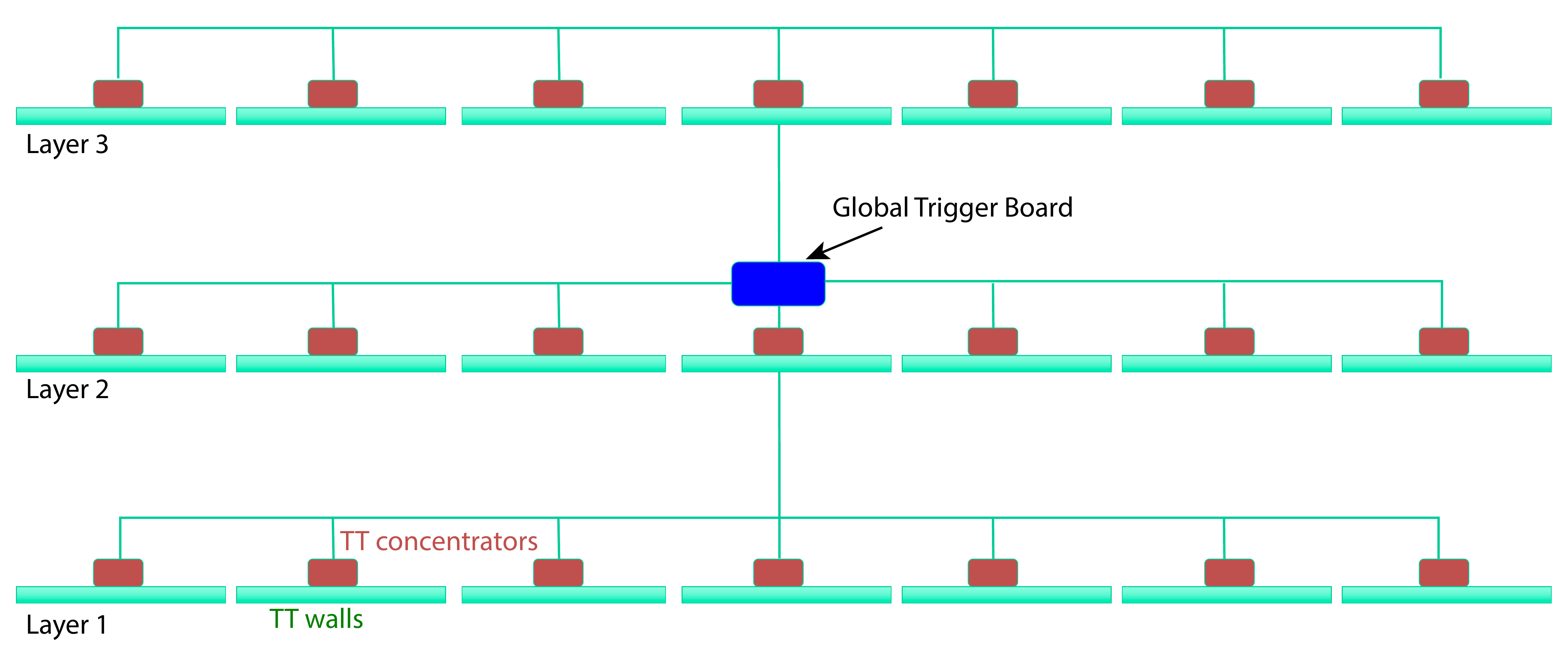}
\caption{Schematic view of the TT global trigger strategy.}
\label{fig:GTB}
\end{figure}

\begin{figure}[hbt]
\centering
\includegraphics[width=\linewidth]{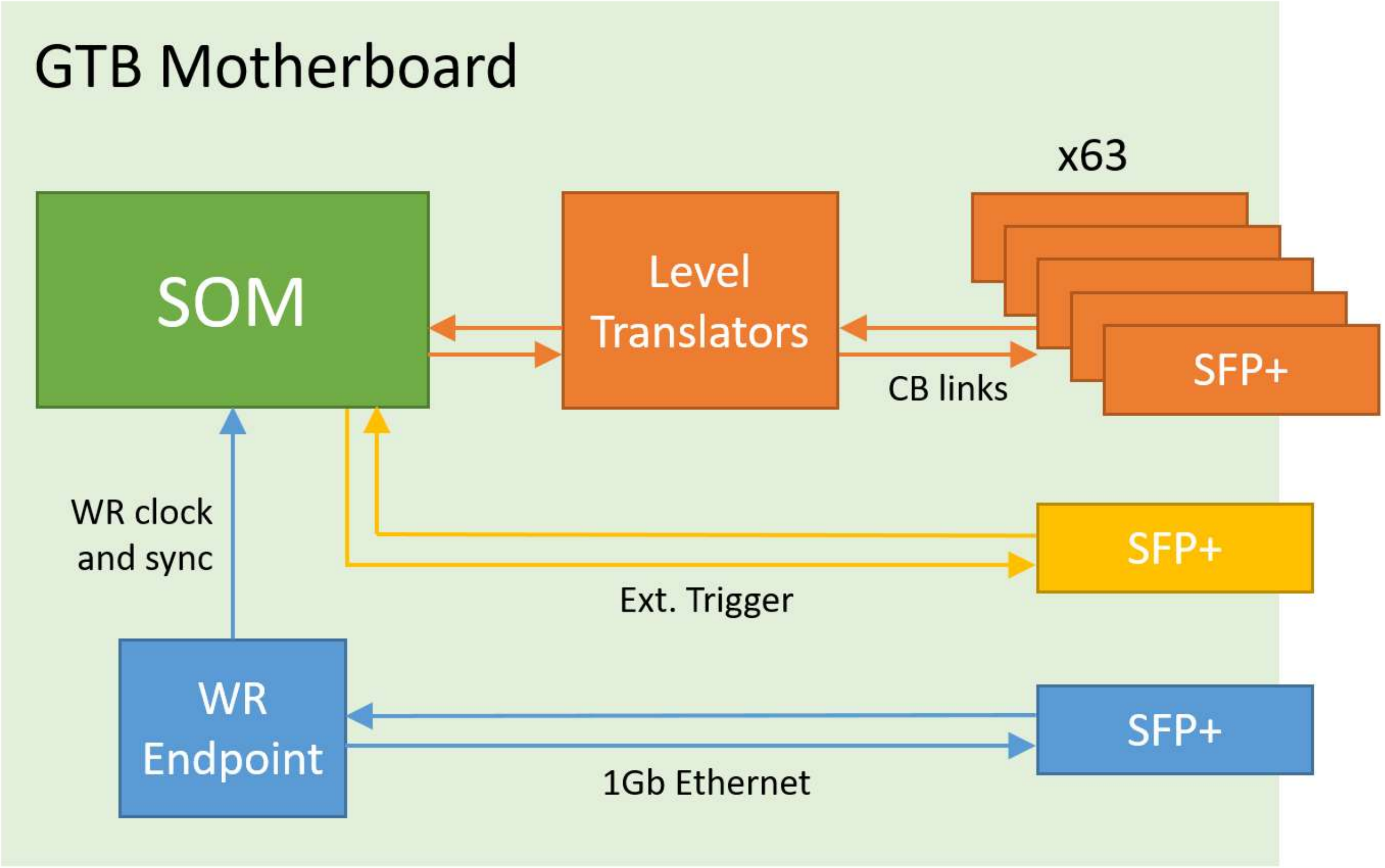}
\caption{Block diagram of the GTB motherboard. The GTB motherboard hosts a SOM, a WR endpoint for clock distribution and synchronisation and 65~bidirectional optical transceivers (SFP+).}
\label{fig:GTB_mother}
\end{figure}

The hardware implementation of the GTB relies on a single device connected to all the CBs (see Section~\ref{cbelectronics}) by means of bidirectional optical links.
Indeed, each CB should send the trigger requests and receive back a validation/rejection that will be used for resetting the FEB and the data acquisition buffers.
Fig.~\ref{fig:GTB} shows a sketch of the connection scheme where the GTB is placed in the middle of the detector.
The GTB should also accommodate a connection to the WR network that would  guarantee the synchronisation with the CBs and will be used for configuring the GTB.
An additional optical link is also needed for the GTB to be able to receive external triggers from the other JUNO systems.
A total of 65~optical links is therefore needed in the GTB as shown in Fig.~\ref{fig:GTB_mother}.
While these two latter links are Gb connections, the 63~links from the CB are not expected to have a bandwidth larger than 400~Mbps since only a partial timestamp will be sent along with the trigger requests.
This is possible because the GTB belongs to the common clock distribution system and hence it receives its full timestamp from the WR network.

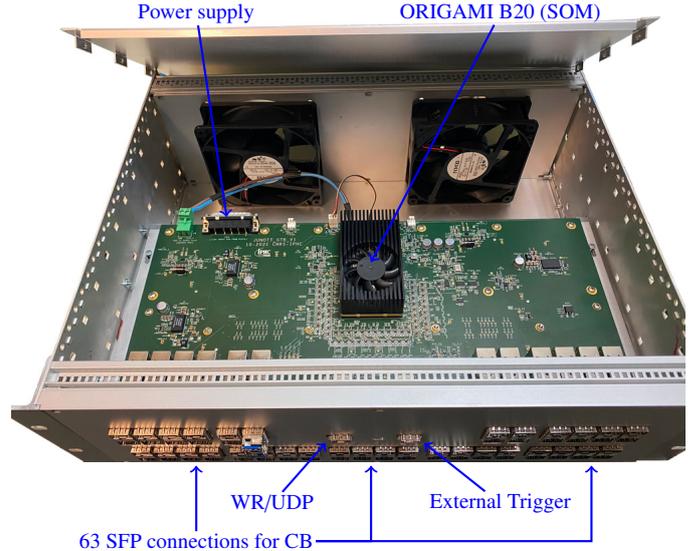
\begin{figure}[hbt]
  \centering
  \begin{tikzpicture}[thick,every node/.style={inner sep=1pt,outer sep=0,color=blue}]
    \node (proto) at (0,0) {\includegraphics[width=\linewidth]{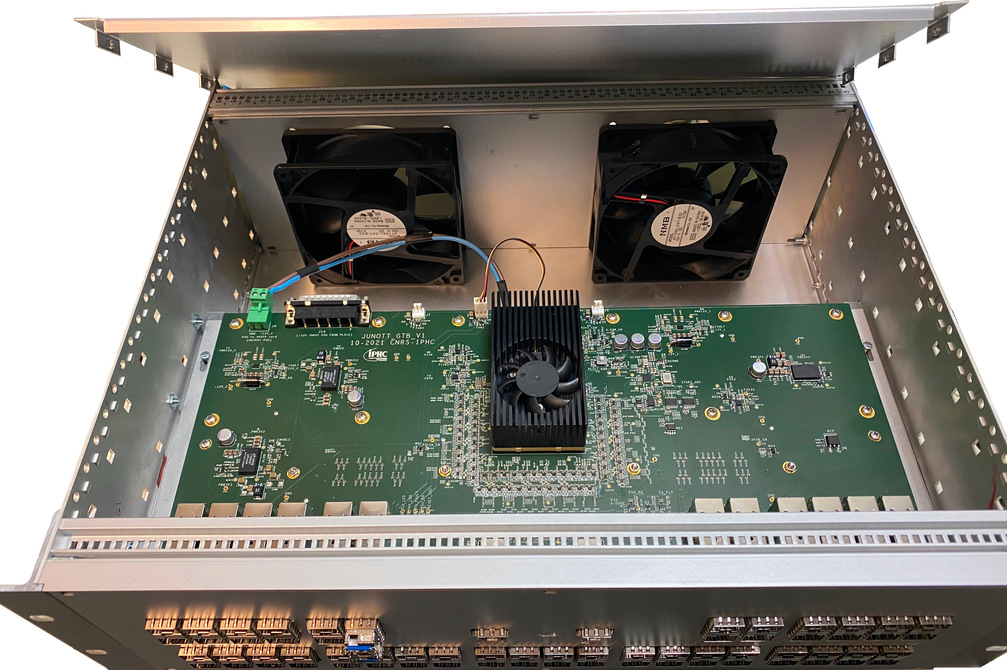}};
    \node (SFPtext) at (-2,-4) {\footnotesize 63 SFP connections for CB};
    \draw[->,blue] (SFPtext) -- (-2,-3) ;
    \draw[->,blue] (SFPtext) -- (3.2,-4) -- (3.2,-3) ;
    \draw[->,blue] (SFPtext) -- (0.3,-4) -- (0.3,-3) ;
    \node (WRtext) at (-1,-3.5) {\footnotesize WR/UDP};
    \draw[->,blue] (WRtext) -- (-0.3,-2.7);
    \node (ExtTrigtext) at (2,-3.5) {\footnotesize External Trigger};
    \draw[->,blue] (ExtTrigtext) -- (1.0,-2.7);
    \node (Origamitext) at (2,3) {\footnotesize ORIGAMI B20 (SOM)};
    \draw[->,blue] (Origamitext) -- (0.3,-0.3);
    \node (PowerSupplytext) at (-2,3) {\footnotesize Power supply};
    \draw[->,blue] (PowerSupplytext) -- (-1.6,0.3);
  \end{tikzpicture}
  \caption{GTB motherboard and SOM.}
  \label{fig:elec:GTB_prototype}
\end{figure}

The present proposal is to implement the GTB as a motherboard hosting a SOM. Fig.~\ref{fig:GTB_mother} shows a block diagram of the GTB motherboard and Fig~\ref{fig:elec:GTB_prototype} shows a photo of the board.
It will accommodate the optical modules and provide the power needed by the SOM that will host an FPGA able to receive the two network links on fast transceivers and the 63~CB links on normal pins in order to
minimise the latency of the link.
For homogeneity with the rest of the system, a Xilinx FPGA has been chosen.
In particular, the commercially available off-the-shelf module Origami~B20~\cite{Origami}
has been chosen as SOM for the GTB.
It hosts a Kintex Ultrascale XCKU060 FPGA~\cite{Xilinx_Kintex_Ultrascale}
that has more than 150 single ended I/Os routed to a Z-Ray connector.

\section{Reconstruction}
\label{sec:reconstruction}

The main goal of the TT reconstruction is to determine the direction of atmospheric muons
crossing the detector.
In order to achieve this goal, the track reconstruction is divided in three successive steps.
In the first step, for each \mbox{MA-PMT}, neighbouring triggered channels are clustered to reduce the impact
of the cross-talk in the reconstruction.
In the second step, the 3D crossing points of the muon are reconstructed based on information from both
perpendicular detection planes of each wall participating in the \mbox{x~--~y} coincidence.
In the final step, a 3D line is fitted to these 3D points via a $\chi^2$ minimisation.
This last step is done independently for every ensemble of more than three 3D points at different
vertical positions, and fits with a bad $\chi^2$ are rejected.

This track reconstruction is able to reduce the remaining radioactive noise rate in the detector
by at least 2 orders of magnitude, as shown in Fig.~\ref{fig:L2_rate}, achieving a rate comparable to
the expected atmospheric muon rate in the TT.
While this algorithm is very efficient to reject background events due to radioactivity, it is also
able to reconstruct almost all muons passing through the detector, with $\sim 97$\%
of the muons having passed the L2 trigger condition being reconstructed in a sample simulated merging
individual muons\footnote{The muons were simulated using the expected energy and angular distribution in the
JUNO cavern.} and the expected radioactive noise in a $\pm15$~\textmu{}s time window around them.

\begin{figure}[htb]
\centering
\includegraphics[width=\linewidth]{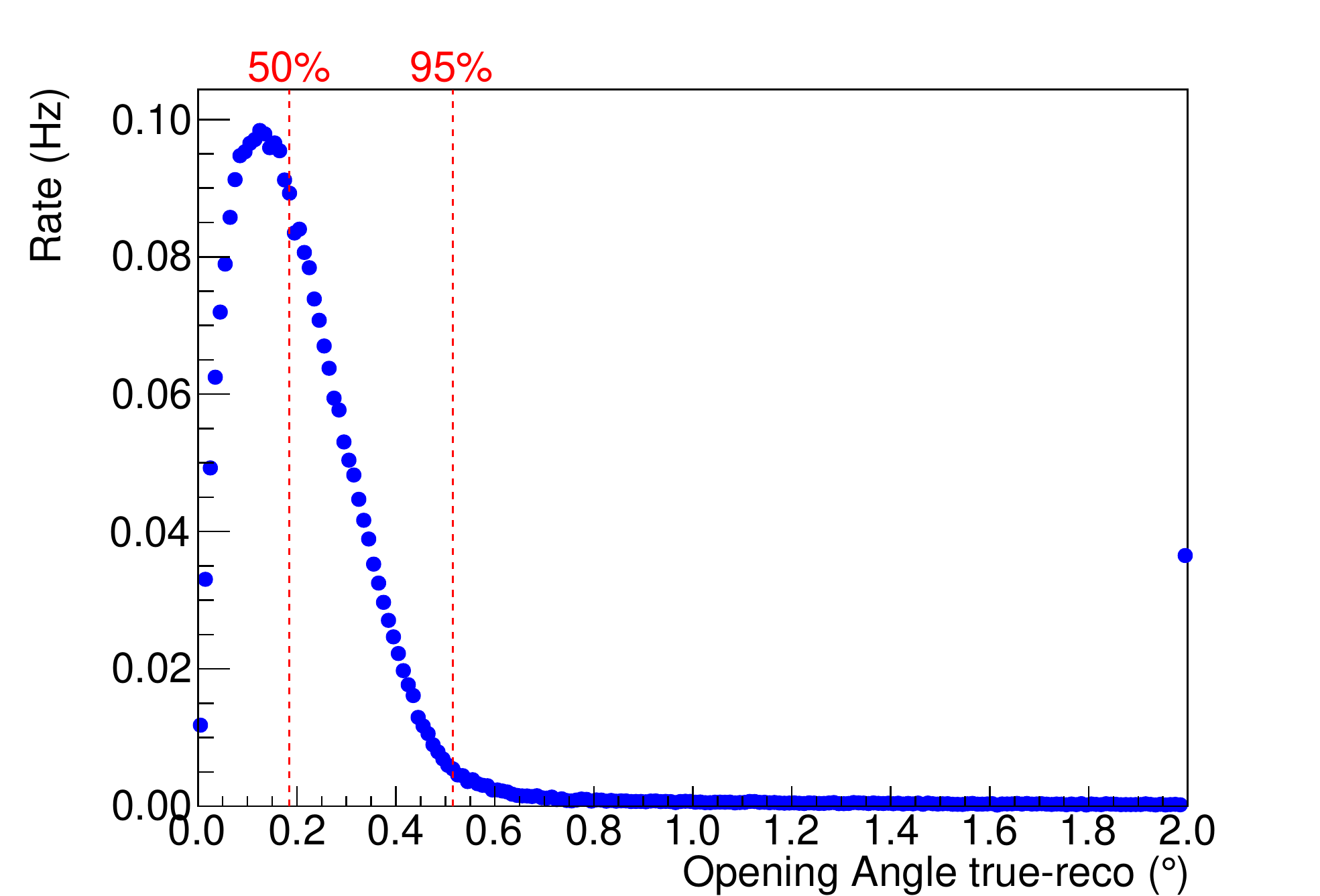}
\caption{\label{fig:rec:angular}Expected angular resolution of the TT reconstruction.
The last point includes also all events having an angular resolution larger than 2$^\circ$.
}
\end{figure}

\begin{figure}[htb]
\centering
\includegraphics[width=\linewidth]{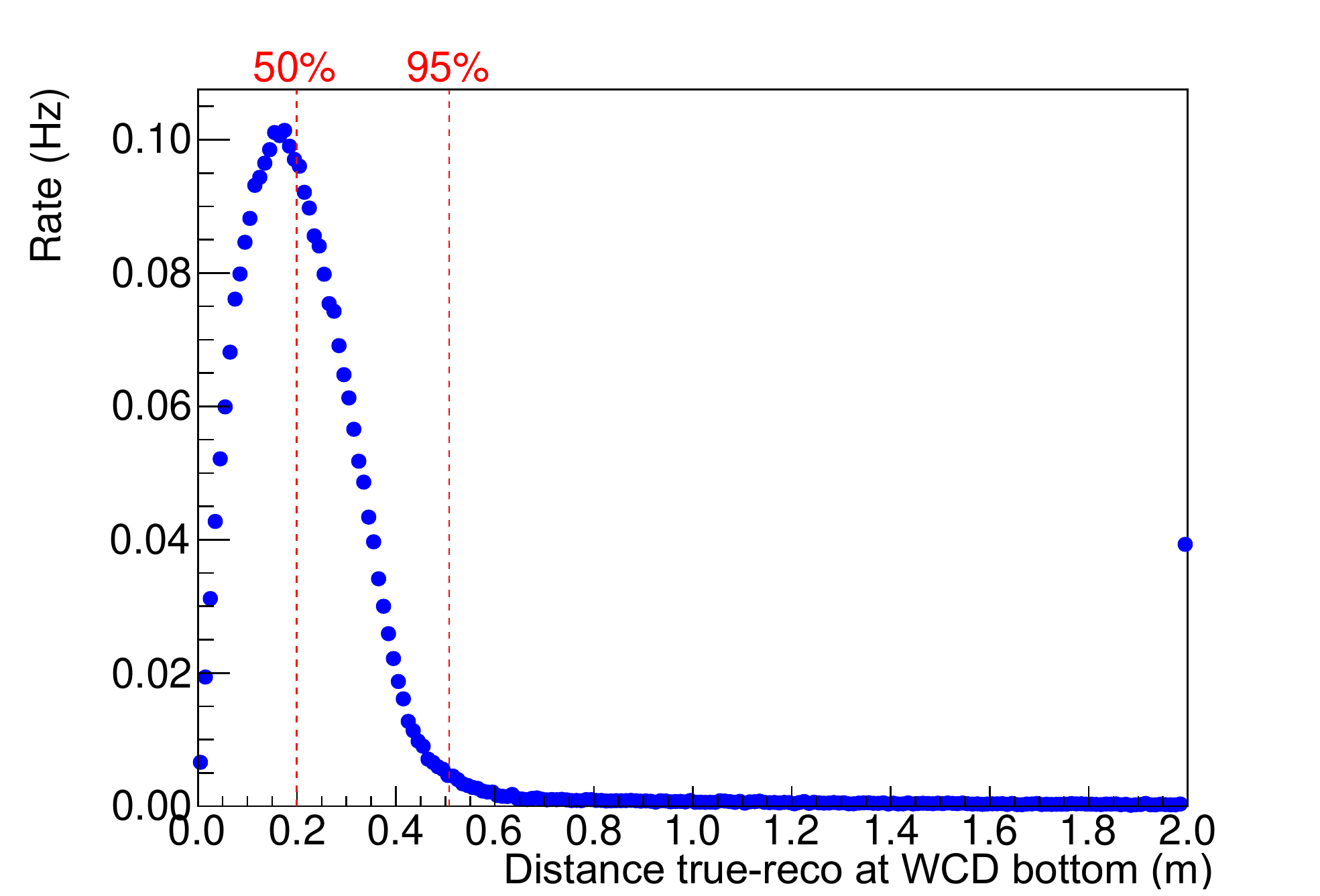}
  \caption{\label{fig:rec:distanceWP}Distance between the points defined by the crossing of the plane
  corresponding to the bottom of the water Cherenkov detector (WCD) and the lines from the
  simulated and reconstructed muons.
  Note that the bottom of the WCD is 45~m below the bottom of the TT, and corresponds to the largest projected error
  of the TT tracking within JUNO.
  The last point includes also all events with a distance larger than 2~m.
}
\end{figure}

Thanks to the high granularity of the TT and the lever arm provided by the distance between its layers,
the median angular resolution of the detector is of about 0.2$^\circ$, as shown in Fig.~\ref{fig:rec:angular}.
Projecting both the simulated and reconstructed tracks to the bottom of the water Cherenkov detector, which
corresponds to the largest error of the reconstruction while still within the JUNO detector, the median
resolution would correspond to about 20~cm, as shown in Fig.~\ref{fig:rec:distanceWP}.

The long tail in these resolutions, containing about 1\% of all reconstructed events, is due to poorly filtered
cross-talk and noise leading to reconstructed tracks not corresponding to the simulated muons.
Some more strict quality criteria for the TT reconstruction
are under evaluation which can significantly reduce this tail, however at the moment
they also reduce by a third the fraction of reconstructed muons, which is not desirable.
Alternative reconstruction algorithms are also under evaluation.
Going beyond looking only at TT data,
a joint reconstruction using the TT, the water Cherenkov detector and the central
detector is expected to be able to easily identify all cases where the TT alone identified a fake muon track and thus
reduce both the long tail in the TT resolution and remove any remaining noise-induced events in
the TT, without negatively impacting on the reconstruction efficiency.
This joint reconstruction is still under study.


\section{TT operation modes} \label{operation}

There are four different modes foreseen for the TT operation.
These modes were thought to make it possible to monitor the TT stability, calibrate it and
perform data acquisition.
Each of those modes can also have additional configurable options for added flexibility, while
keeping the same global goal.
The four modes foreseen, which will be described in details in the following subsections,
are the trigger rate test (TRT), pedestal measurement (PED), LED light injection (LED), and normal data taking
modes.
Tab.~\ref{tab:operation:modes} shows some characteristics for these four modes.

\begin{table}[htb]
  \centering
  \caption{TT operation mode characteristics}
  \label{tab:operation:modes}
    \begin{tabular}{l|cccc}
      & TRT      & PED      & LED      & Normal   \\ \hline
      Trigger            & auto     & external & external & auto     \\
      Charge measured    & no       & yes      & yes      & optional \\
      Zero-suppression   & --       & no       & no       & optional \\
      Fixed data size    & yes      & yes      & yes      & no \\
    \end{tabular}
\end{table}

\subsection{Trigger rate test mode} \label{TRTmode}

This mode is used to measure the trigger rate per channel.
The main goals of this mode are to characterise the electronic noise,
define the trigger threshold and search for light leaks in the TT modules.
In the TRT mode, the acquisition is set in auto-trigger mode without charge acquisition, only the number of triggers in a given time window (typically 1~s)
for each channel is returned.
It is also foreseen that the trigger rate of all channels will be monitored during data taking automatically, in addition to this acquisition mode.
The main advantage of having the possibility to request this data
is to reduce the latency in this measurement for direct usage during the
above-mentioned calibration and search for light leaks.

\begin{figure}[htb]
  \centering
\includegraphics[width=\linewidth]{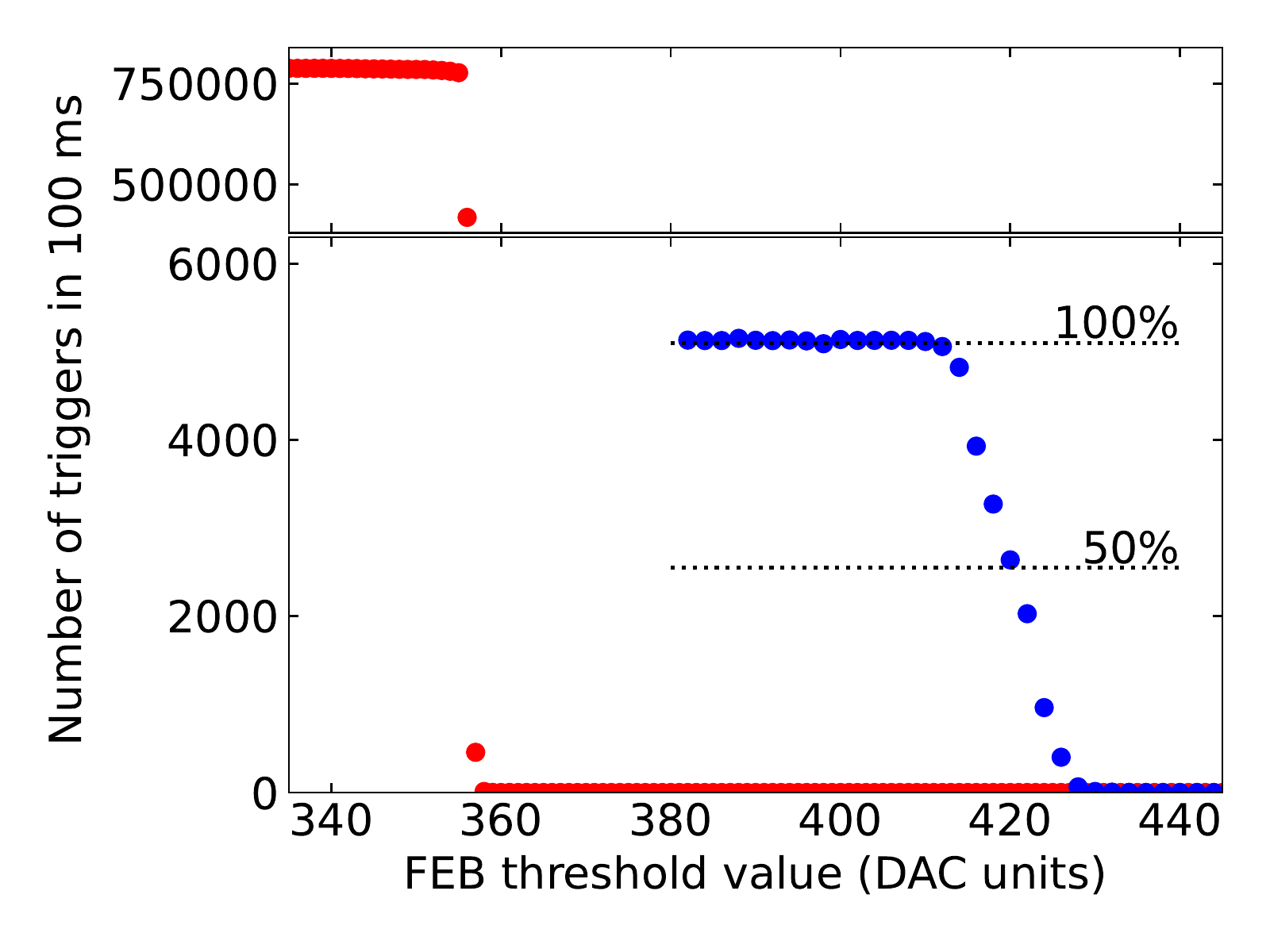}
  \caption{TRT measurement as a function of the threshold set in one channel of MAROC3
  with no charge injected (red) and with $\approx 53$~fC injected at 50~kHz (blue).
  }
  \label{fig:operation:TRT:scurve}
\end{figure}

Fig.~\ref{fig:operation:TRT:scurve} shows an example of the determination of the
threshold and electronic noise for one particular channel of the FEB using multiple acquisitions
using the TRT mode with a varying threshold set in the FEB.
For the determination of the electronic noise no charge is injected, while
for the determination of the threshold a known charge (corresponding to 1/3~p.e.) is injected in the tested channel.
Both measurements have been done, using a test setup, for all channels of all FEB cards to validate
the FEB production.
In all accepted cards, a clear distinction between the electronic noise (red)
and the 1/3~p.e.\@{} injected charge (blue)
is observed,
with a change of the 50\% threshold from 357~DAC units to 420~DAC units in Fig.~\ref{fig:operation:TRT:scurve}.
The exact thresholds
are not at the same DAC values for each FEB and have to be set individually from the values obtained during testing.
For all accepted FEB, the standard deviation of threshold values among the 64 channels in the board
is in average 6.5~DAC units, and it is always below 10~DAC units.

\begin{figure}[htb]
  \centering
\includegraphics[width=\linewidth]{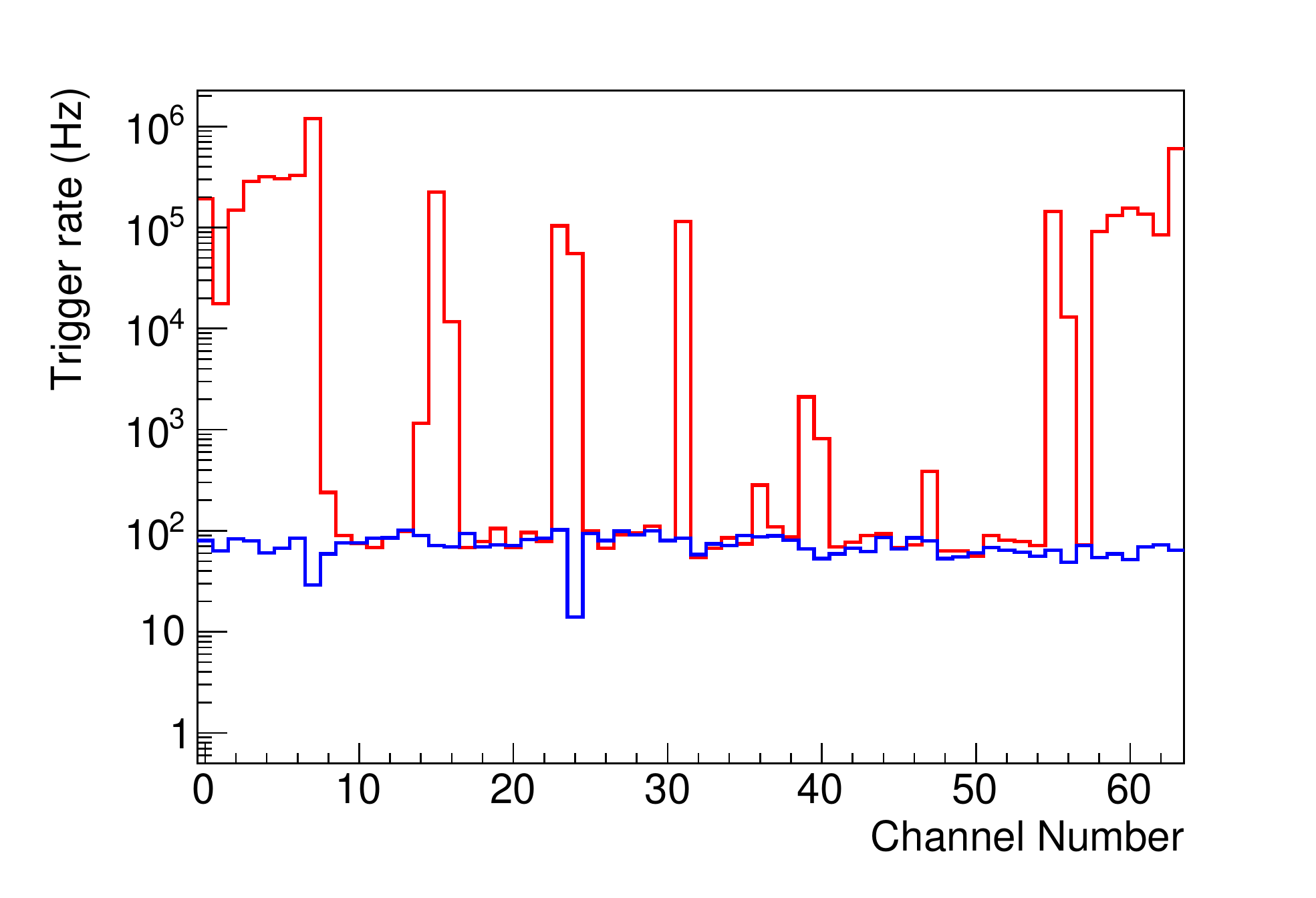}
    \caption{Trigger frequency versus channel for one sensor before (red) and after (blue) fixing
    noisy channels for one of the cards of the TT prototype.}
    \label{fig:operation:TRT:channel}
\end{figure}

In addition to determining the desired operational threshold, the TRT mode
can be used to check noisy channels during normal operation.
The presence of noisy channels can be induced, for example, by light leaks inside the detector.
Fig.~\ref{fig:operation:TRT:channel} shows an example of such a measurement
done on the TT prototype (described in Sec.~\ref{ttproto}) before and after fixing
a grounding issue due to an improper installation of the ROB.

\subsection{Pedestal mode} \label{pedestalmode}

\begin{figure}[hbt]
  \centering
\includegraphics[width=\linewidth]{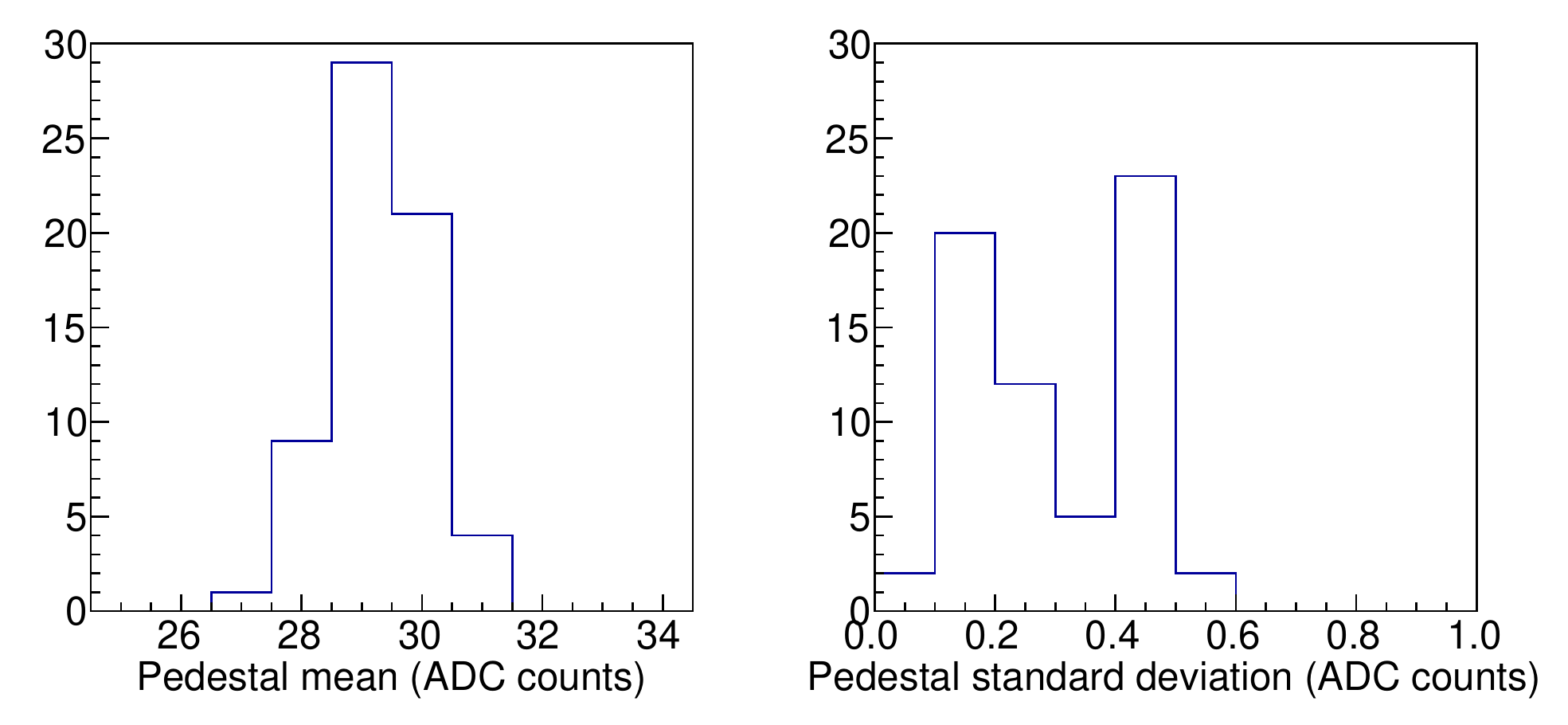}
    \caption{Mean (left) and standard deviation (right) of pedestal data for one \mbox{MA-PMT} used in the TT prototype. Wilkinson charge readout used.}
    \label{fig:operation:PED}
\end{figure}

This mode is used to determine the measured charge in the absence of a signal.
This is achieved by measuring the charge with a fixed trigger frequency rather than using
the \mbox{MA-PMT} signal to trigger the detection.
It is possible that some of the fixed triggers will coincide in time with regular triggers from the \mbox{MA-PMT}
by chance.
These rare coincidences will produce events with charge away from the pedestal peak and
can easily be removed from the pedestal distribution when the pedestal run is being analysed.
Fig.~\ref{fig:operation:PED} presents the pedestal values of the 64 channels of one sensor (left)
and their standard deviation values (right).
The mean of the pedestal standard deviation,
using the 8-bit Wilkinson charge readout mode,
is of the order of 0.3~ADC counts, representing about 0.04~p.e.

\subsection{LED light injection mode} \label{LEDmode}

This mode works similarly to the Pedestal mode described before, however the measurement
happens at about the same time of the light injection.
As in the Pedestal mode, in the LED mode data is taken with a fixed trigger frequency, with which the
LED is pulsed ignoring the \mbox{MA-PMT} trigger.

\begin{figure}[b]
  \centering
    \includegraphics[width=\linewidth]{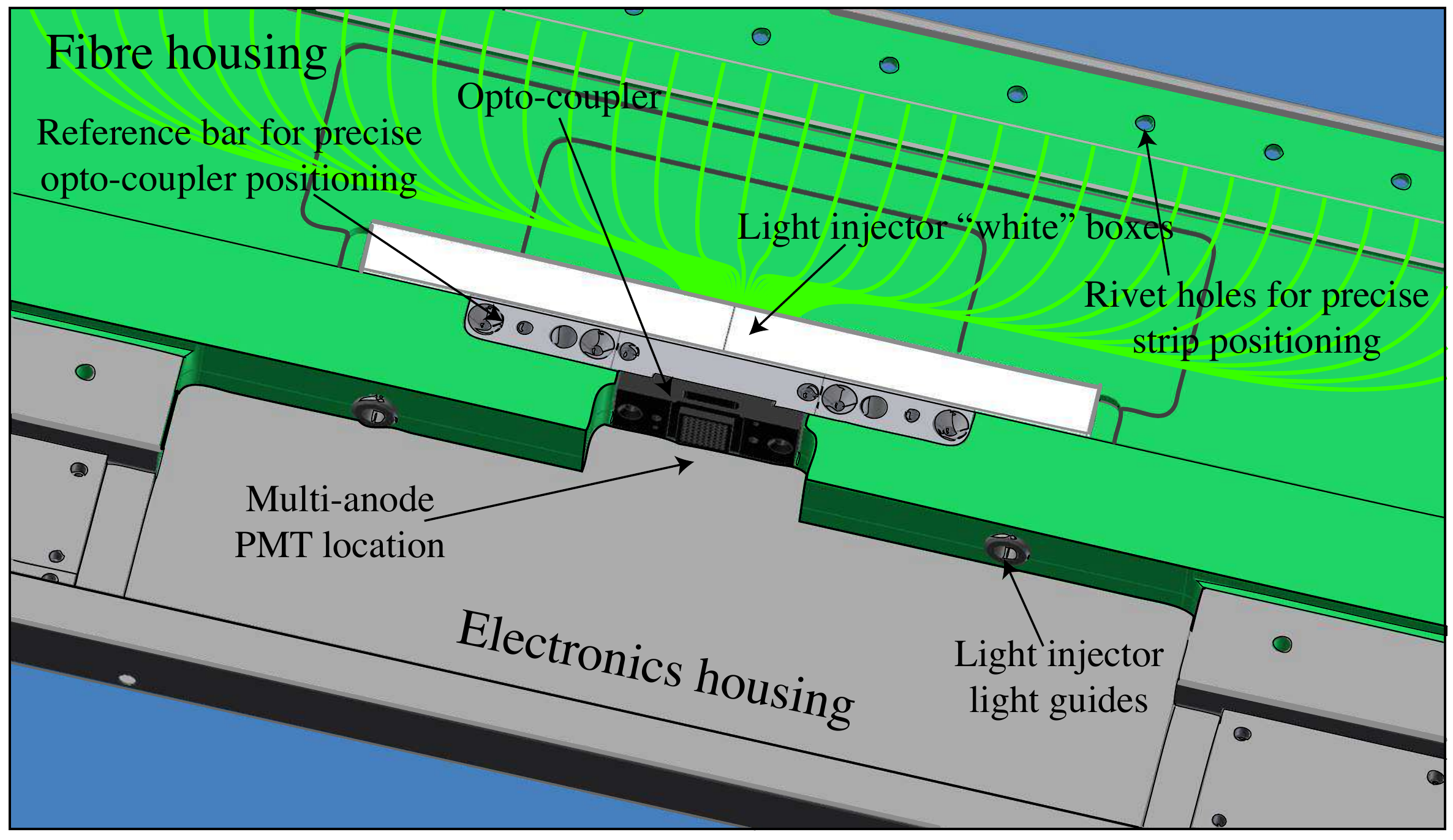}
    \caption{3D view of the central part of an end-cap.}
    \label{endcap_3D}
\end{figure}

The light injected in the system uses the same system described
in~\cite{Adam:2007ex}.
In this system, light is injected at each end-cap into the WLS fibres just in front of the \mbox{fibre--MA-PMT}
opto-coupler with the help of two blue LEDs, straight PMMA light
guides and a white painted diffusive box as shown in Fig.~\ref{endcap_3D}.

\begin{figure}[htb]
  \centering
\includegraphics[width=\linewidth]{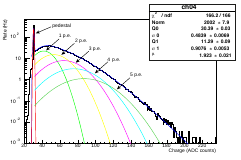}
    \caption{Charge distribution using the light injection system.}
    \label{fig:operation:LED:distribution}
\end{figure}

The LEDs are pulsed by a signal from the ROB, which is equipped with a driver designed for this
application, as described previously.
After a certain configurable delay from the LED pulse, to account for the time required to flash the LED,
an external trigger is generated to perform the charge measurement in the MAROC3.
The LED intensity can be configured to produce a signal with a mean intensity around the single
p.e.\@{} as shown in Fig.~\ref{fig:operation:LED:distribution} with the digitized charge readout
for the 8-bit Wilkinson mode.
In this figure a fit to the SPE distribution using the function described
in~\cite{Bellamy:1994bv} is also drawn.
In this distribution the peak related to the pedestal is still clearly visible, with an additional
set of distributions corresponding to the cases where one to five p.e.\@{}
have been detected by that \mbox{MA-PMT} channel.
In this case, the measured mean intensity of the light ($\mu$) corresponds to
${1.92 \pm 0.02}$~p.e.\@{} and the
1~p.e.\@{} measured charge ($Q1$) is ${11.29 \pm 0.09}$~ADC units.
This measured charge is used to calibrate the conversion between ADC units and charge as
the PMT gain had been set to $10^6$.

\begin{figure}[b]
  \centering
\includegraphics[width=\linewidth]{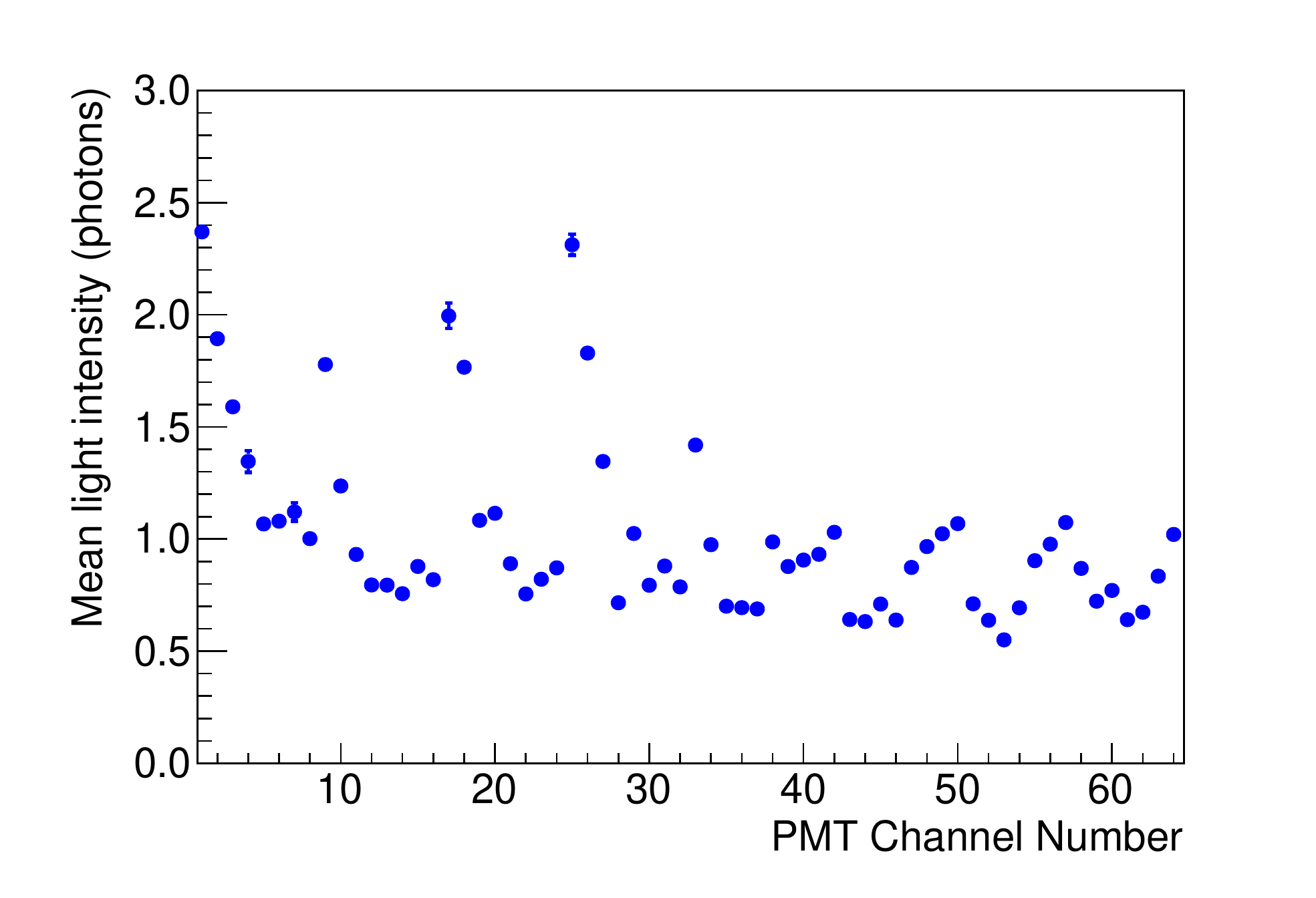}
    \caption{Mean light intensity per channel using the light injection system for one \mbox{MA-PMT}.
    Variations of the mean intensity are due to geometrical effects.
    }
    \label{fig:operation:LED:fit_mu}
\end{figure}

While the system is designed to produce an approximately uniform illumination of all fibres, since the illumination is done from the sides, the fibres
at the edge of the opto-coupler receive more light and shadow the more central fibres.
Due to these geometrical effects, the illumination in the different fibres might vary
by a factor of 3, as shown in Fig.~\ref{fig:operation:LED:fit_mu}.
For an optimal performance of the fit described before, it is suitable to tune the light
intensity to have the average number of photons observed in the 1--2~p.e.\@{} range~\cite{Anfimov:2018qjz}.
In this range, both the pedestal
and the single photo-electron peak are clearly visible.

\begin{figure}[htb]
  \centering
\includegraphics[width=\linewidth]{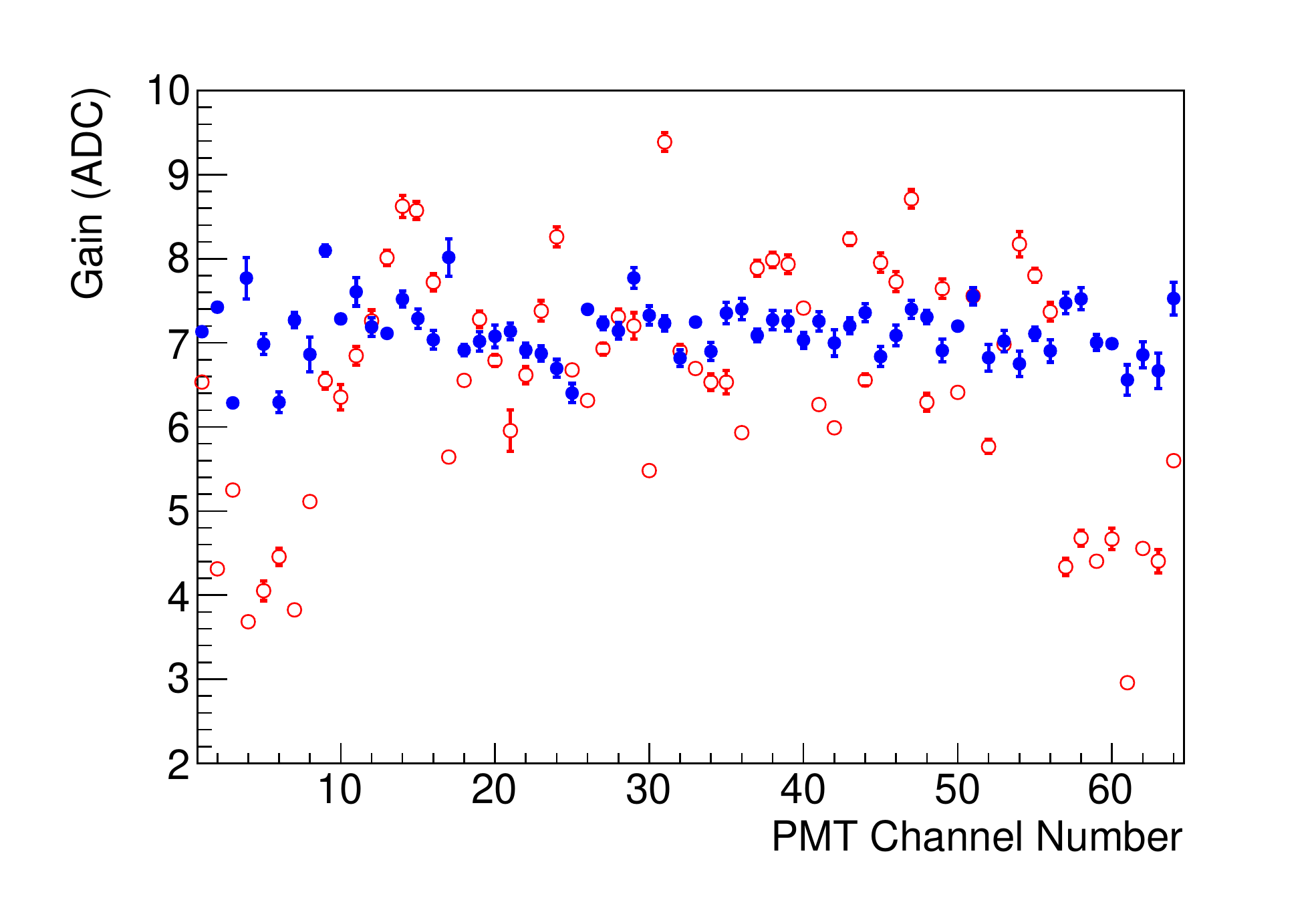}
    \caption{Gain per channel measured using the light injection system.
    Red empty (blue full) markers show the gain before (after) the gain equalisation procedure.
    }
    \label{fig:operation:LED:fit_gain}
\end{figure}

The light injection system can then be used to define the g.e.f.\@{} to be
used in the MAROC3 to equalise the gain in all channels of the \mbox{MA-PMT}.
The measured gain in each channel before (red)
and after (blue) gain equalisation using these g.e.f.\@{}
are shown in Fig.~\ref{fig:operation:LED:fit_gain}.
Before equalisation, gains vary by a factor of up to 3, while after this procedure gains
vary only by 30\%.

This LED mode is foreseen to be used regularly during the whole duration of the experiment,
mainly to monitor the \mbox{MA-PMT} gain.

\subsection{Normal run mode} \label{normalmode}

This mode is the one used for regular data acquisition.
In this case the \mbox{MA-PMT} signal passing the threshold will produce a MAROC3 trigger
as described beforehand.
Additionally, zero-suppression in the ROB can be enabled in this mode to reduce the data size.

To safeguard against unexpectedly high rates, it is also possible to entirely disable
the charge readout, which reduces the detector dead-time and
significantly decreases the data size, while keeping the information about
the triggered channels.
A mixed acquisition option with and without charge readout
is being investigated.
In this case, a new trigger will start charge readout if the acquisition system is not already
performing the charge readout for another trigger.
However, in the case where the acquisition system
is busy, no charge readout will be performed and only the triggered channels and trigger time will be saved.

The data produced by the TT readout system in the normal mode contains the \mbox{MA-PMT} trigger time-stamp provided by the CB,
the trigger register identifying which channels on the \mbox{MA-PMT} triggered, and the measured charges
if charge readout is enabled.


\section{The TT prototype} \label{ttproto}

In order to test the electronics under development, a prototype of the TT was built at
IPHC\footnote{Institut Pluridisciplinaire Hubert {\scshape Curien}, Strasbourg, France.}
using the worst modules produced for the OPERA experiment.
It is also equipped with the same electronics cards as the TT for testing.

\begin{figure}[htb]
\centering
    \includegraphics[width=.95\linewidth]{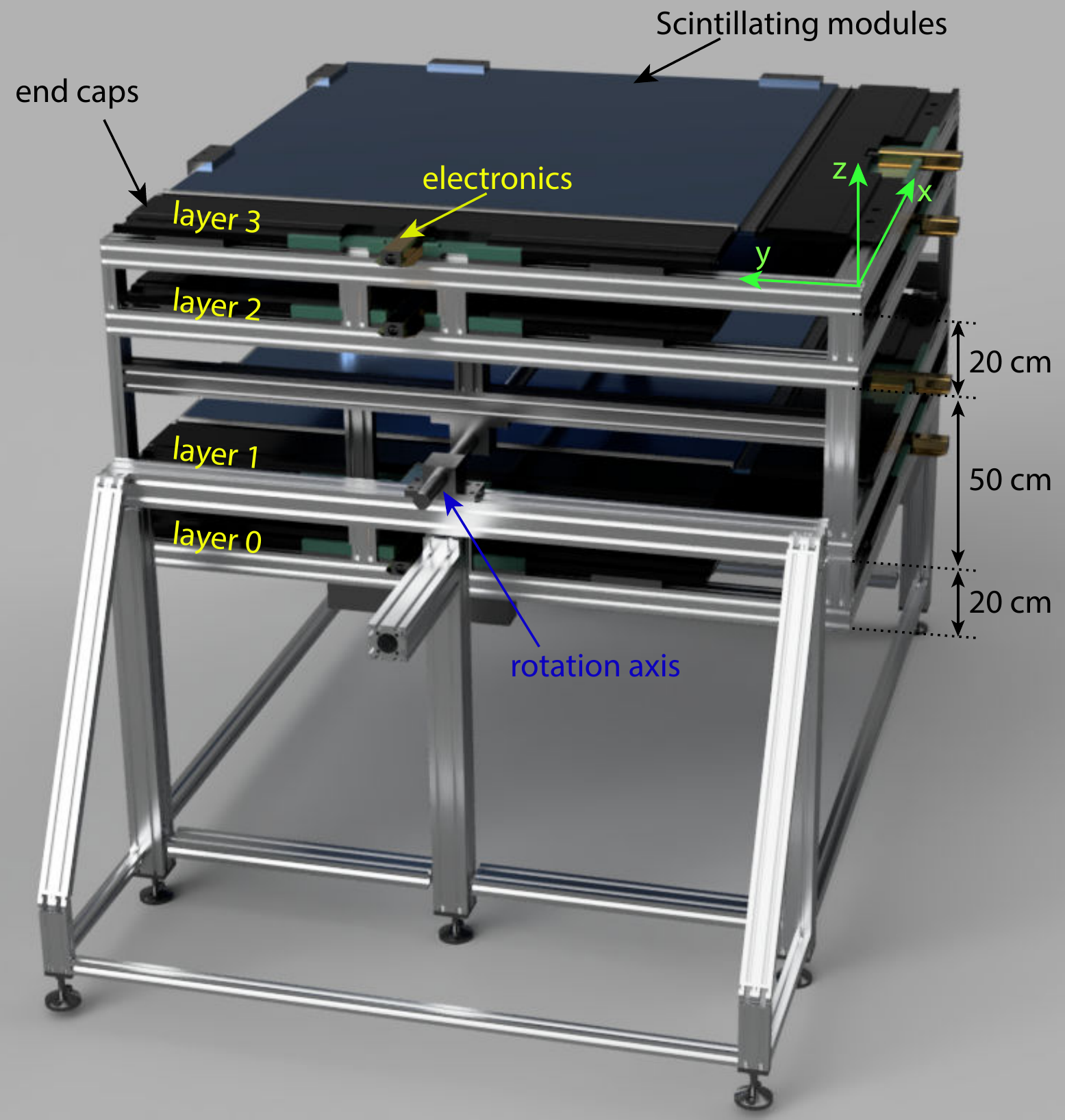}
    \caption{Schematic view of TT prototype.}
    \label{fig:ttproto:drawing}
\end{figure}

This prototype, shown in Fig.~\ref{fig:ttproto:drawing}, is made of 4~layers.
As in the case of the TT, each layer is composed of two planes oriented perpendicularly.
Each plane on the prototype was built by cutting a TT module in 3 parts at a distance of
about 1.7~m from each end, with the left and right end-caps becoming the x and y oriented planes,
respectively, which are used for the \mbox{x~--~y} coincidences as in the TT.
This means that each layer of the prototype was made from a single TT module and
has a surface of about $1/16$ of the surface of one TT wall.
Differently from the TT modules which are read on both sides, the prototype
ones are only read on one side.

\begin{sloppypar} 
In each of the 8~modules end-caps was installed one \mbox{MA-PMT}, one FEB, and one ROB.
All the 8~ROBs are connected to one CB placed in the metal frame supporting the prototype.
In the prototype it is also possible, for testing purposes, to read the data directly from the ROB
without passing by the CB, by using a different version of the ROB specifically designed for that purpose.
Using the prototype, it has already been possible to validate the calibration procedures
and test the interface between the various electronics cards using natural signals,
albeit with smaller rate than the one expected in JUNO,
due to the lack of the high rate of events expected from the rock radioactivity.
It is worth noting that most figures in Sec.~\ref{operation} were obtained from
data of this prototype.
\end{sloppypar}

\begin{figure}[htb]
\centering
\includegraphics[width=\linewidth]{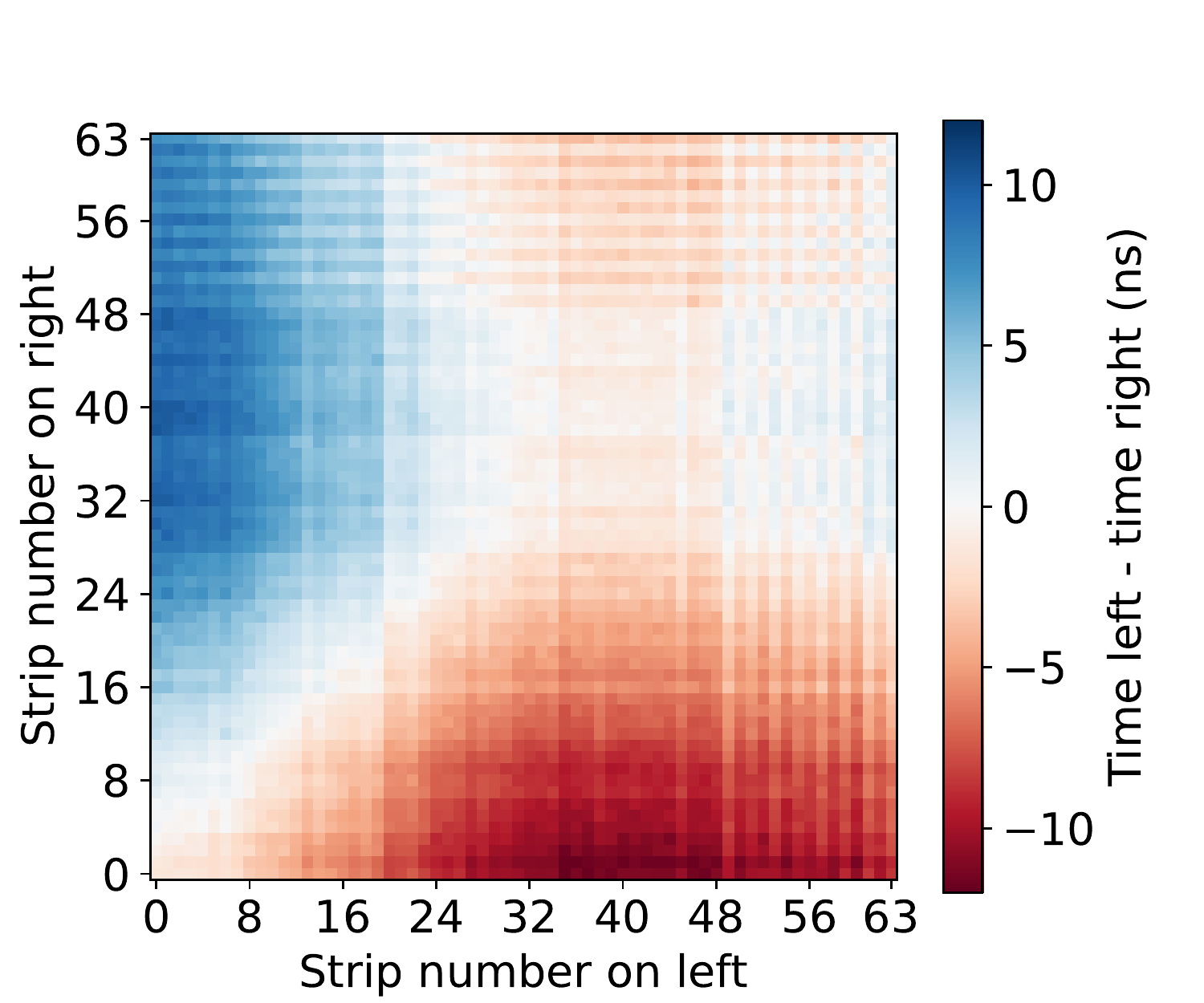}
    \caption{Distribution of the average time difference between signals in left and right cards as
    a function of the strip which triggered the \mbox{MA-PMT} in each card during `normal' data taking.}
    \label{fig:ttproto:DeltaT_2D}
\end{figure}

Using the TT prototype
several consecutive days of data have been collected.
This data makes it possible to check, e.g., the difference in the coincidence time between
the two planes of the same layer, as shown in Fig.~\ref{fig:ttproto:DeltaT_2D}.
In this figure the average time difference between the \mbox{MA-PMT} trigger on the left and right
cards is shown as a function of the channel triggered in each card.
This distribution reflects the different length of the fibres from each interaction point
to the \mbox{MA-PMT} for the x and y directions.

\begin{figure}[htb]
  \centering
\includegraphics[width=\linewidth]{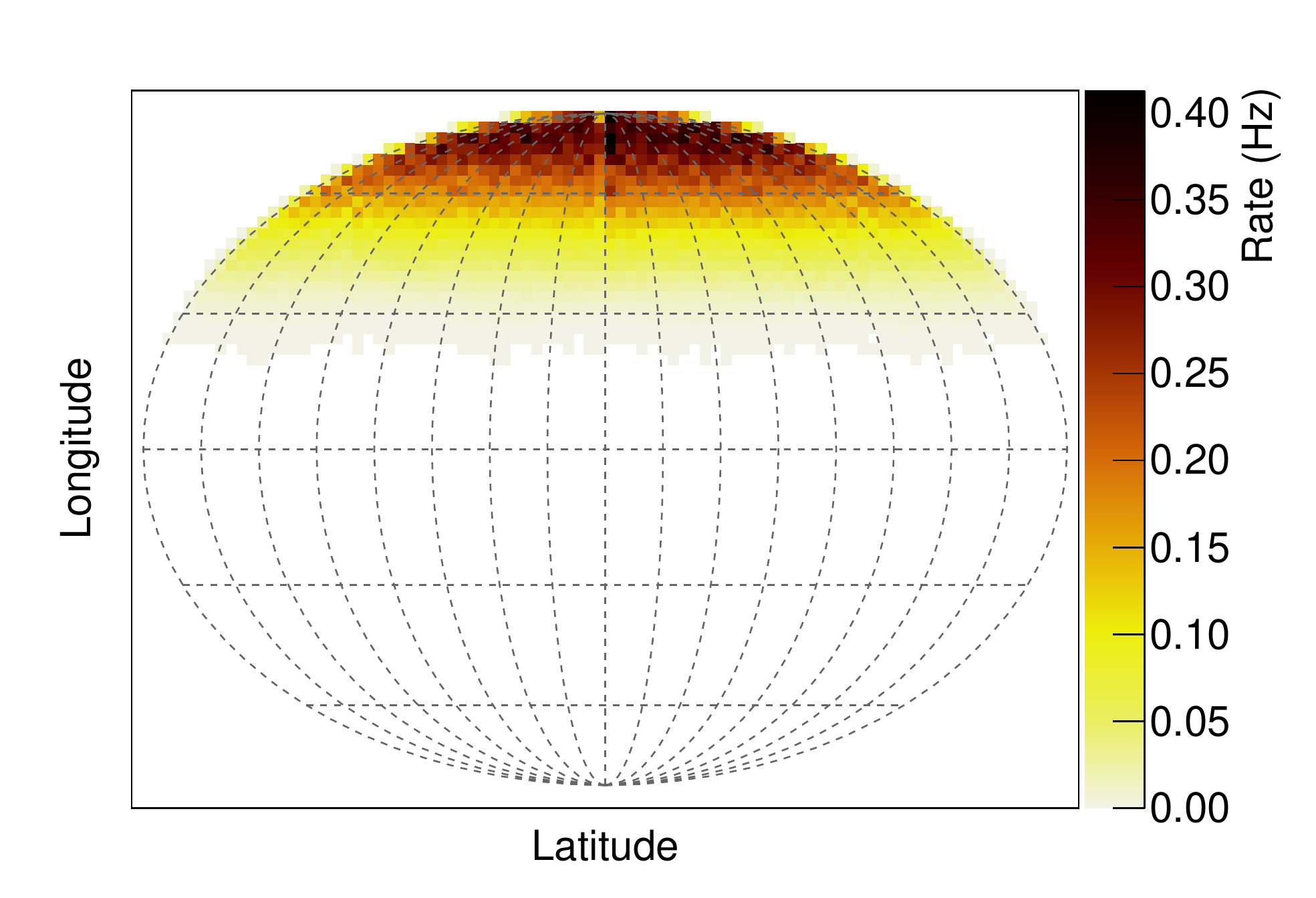}
    \caption{Distribution of the direction of muons observed in the prototype.
    The 3D sphere is projected in a plane using the Mollweide projection.
    Longitude lines are shown every $30^\circ$,
    while latitude lines are shown every $22.5^\circ$.
    The prototype is facing upwards during this data taking.
    Colour code corresponds to the observed muon rate.
    }
    \label{fig:ttproto:skymap}
\end{figure}

Using the data from the TT prototype it is possible to measure the atmospheric muon flux distribution
at IPHC by requiring coincidences between 3 or 4 layers of the prototype,
and reconstructing the muon tracks.
Fig.~\ref{fig:ttproto:skymap} shows the reconstructed muon directions
obtained using the Mollweide projection.
Given the requirement of 3 or 4 layers in coincidence for the reconstruction,
the field of view of the telescope is limited to about 70$^\circ$ from the vertical.
A motor is added to the prototype rotation axis as shown in Fig.~\ref{fig:ttproto:drawing}
to increase this field of view by rotating the whole device.

\begin{figure}[htb]
  \centering
\includegraphics[width=\linewidth]{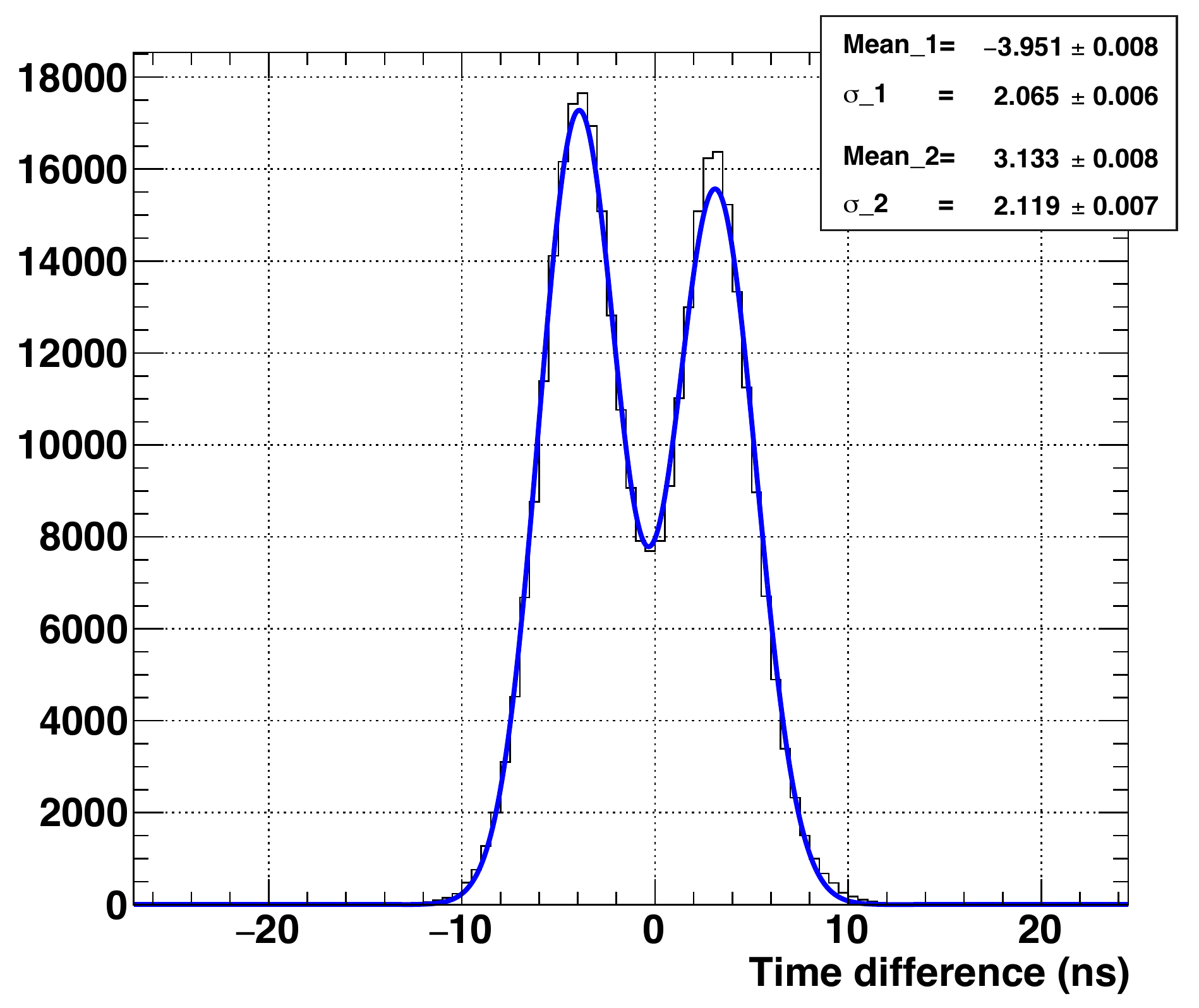}
    \caption{Time difference between the first and last layers of the prototype from atmospheric muons when the prototype is in vertical position.
    }
     \label{fig:timing}
\end{figure}

In order to have a good estimate of the timing capabilities the prototype has been placed in vertical position
(i.e., a 90$^\circ$ rotation with respect to the position shown
in Fig.~\ref{fig:ttproto:drawing}) to record horizontal atmospheric muons.
In this position, it is expected to record about the same number of muons coming from each side.
Fig.~\ref{fig:timing} presents the time difference between hits recorded on first and last layers of the prototype.
The distance between these two layers being of 90~cm it is expected to observe a time difference of the order of 3~ns assuming all muons travel at the speed of light.
The two peaks observed on Fig.~\ref{fig:timing} correspond to muons coming from either side of the detector.
They are centred at about $\pm$3~ns, with 0 corresponding to events being seen at the same time in both layers.
The time dispersion is of the order of 2~ns and includes the decay time in the scintillator (2.3~ns) and the WLS fibres (7.6~ns), and the time-walk in the electronics threshold.
All other delays, as the difference in length of fibres from one channel to another, have been corrected.
The time-bin of the clock used for timestamping triggers during this data taking was of 1.04~ns.
The slight asymmetry between the two peaks is probably due to the other objects around the prototype.

\section{Installation} \label{installation}

The 496~TT modules were transported from LNGS to the JUNO site in 2017.
According to the schedule, the installation of the TT will begin after the assembly of the central detector, presumably end of 2023.

A dedicated  support structure providing the necessary rigidity has been designed (Fig.~\ref{ttjuno}).
It includes two main beams of 48~m across the experimental hall, 3~layers of the transverse girders of 20~m between them, a support of the JUNO calibration system, and the support frames of the TT walls.
 
\begin{figure}[hb]
\centering
\includegraphics[width=.9\linewidth]{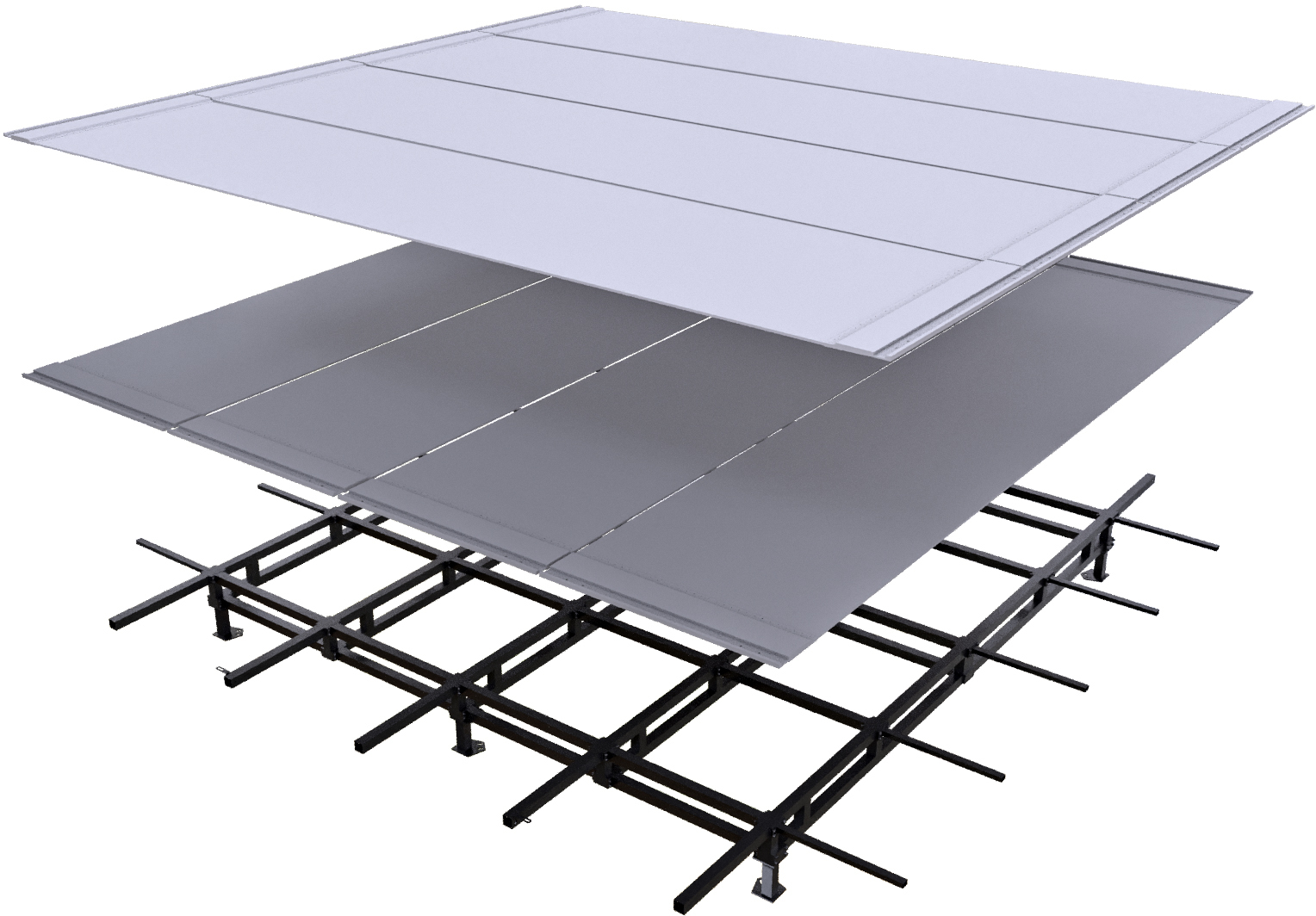}
  \caption{The TT wall, measuring ${7.6 \times 7.6}$~m$^2$, assembly on its support frame called a ``table''.}
\label{Table}
\end{figure}

The walls will be assembled on their support frames, called ``tables''.
4~modules will be placed on the table in one direction (x) and other 4 modules in the perpendicular direction (y), as shown in Fig.~\ref{Table}.

\begin{figure}[ht]
\centering
\includegraphics[width=.9\linewidth]{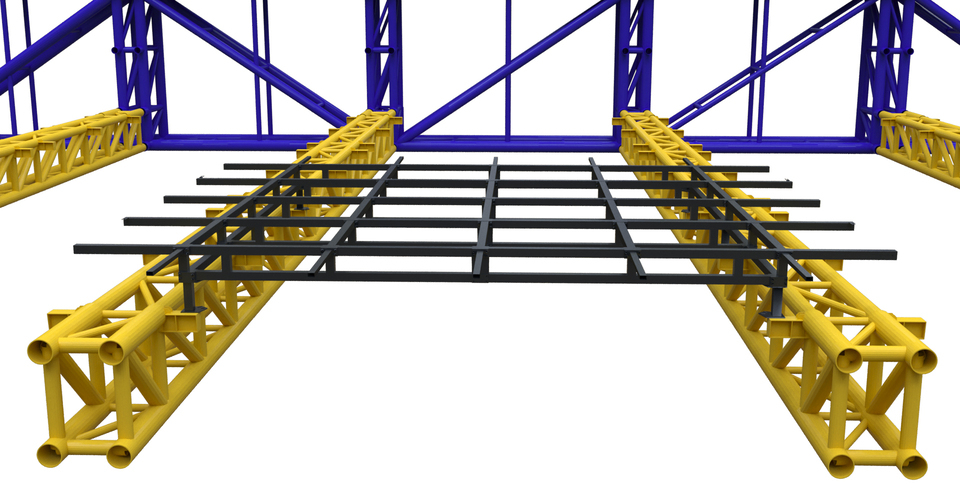}
\caption{A scheme of the installation of the TT ``tables'' in the TT support.}
\label{TableInJUNO}
\end{figure}

All the modules will be equipped with the FEB and ROB and commissioned before installation.
Then, each wall will be equipped with the CB and with the cables to connect all  ROBs to the CB.
Assembled this way, the walls will be commissioned (the counting rate of each \mbox{MA-PMT} should be similar and corresponding to the local atmospheric muons and radioactivity rate) and then moved by a crane to their final place in the detector. The tables are rigid enough to keep the walls flat during the manipulations by the crane and in the detector after the installation  where they rest on the transverse girders, as shown in Fig.~\ref{TableInJUNO}.

In total, 60~walls will be installed in 3~layers separated by 150~cm in height.
To avoid  inefficiency  in the detection of atmospheric muons crossing regions between neighbouring  walls, the walls overlap by 150~mm on all sides.
For that, the overlapping walls have slightly different height of their position and have to be installed in a proper order.
In addition to that, the ``tables'' of the three walls above the chimney have a different design providing a separation by only 21~cm in their vertical position.
During the installation, the position of the walls will be measured by survey instruments with an accuracy of about 1~mm.
The relative alignment of the walls can be further verified with the help of atmospheric muons during the data taking.

According to the installation schedule, one wall per day will be assembled and mounted in the detector.
This time includes some time for curing the modules of any light leaks that might have appeared due to
transportation or storage conditions.

\section{Ageing monitoring}
\label{sec:ageing}

\begin{sloppypar} 
The production of the TT strips have been done by \mbox{AMCRYS-H} company~\cite{amcrys} during two years from 2004 to 2006. It is expected that some ageing of the plastic scintillator will occur before the TT will be used in JUNO experiment, and will continue during the whole lifetime of the detector.
\end{sloppypar}

\begin{figure}[htb]
\centering
\includegraphics[width=.8\linewidth]{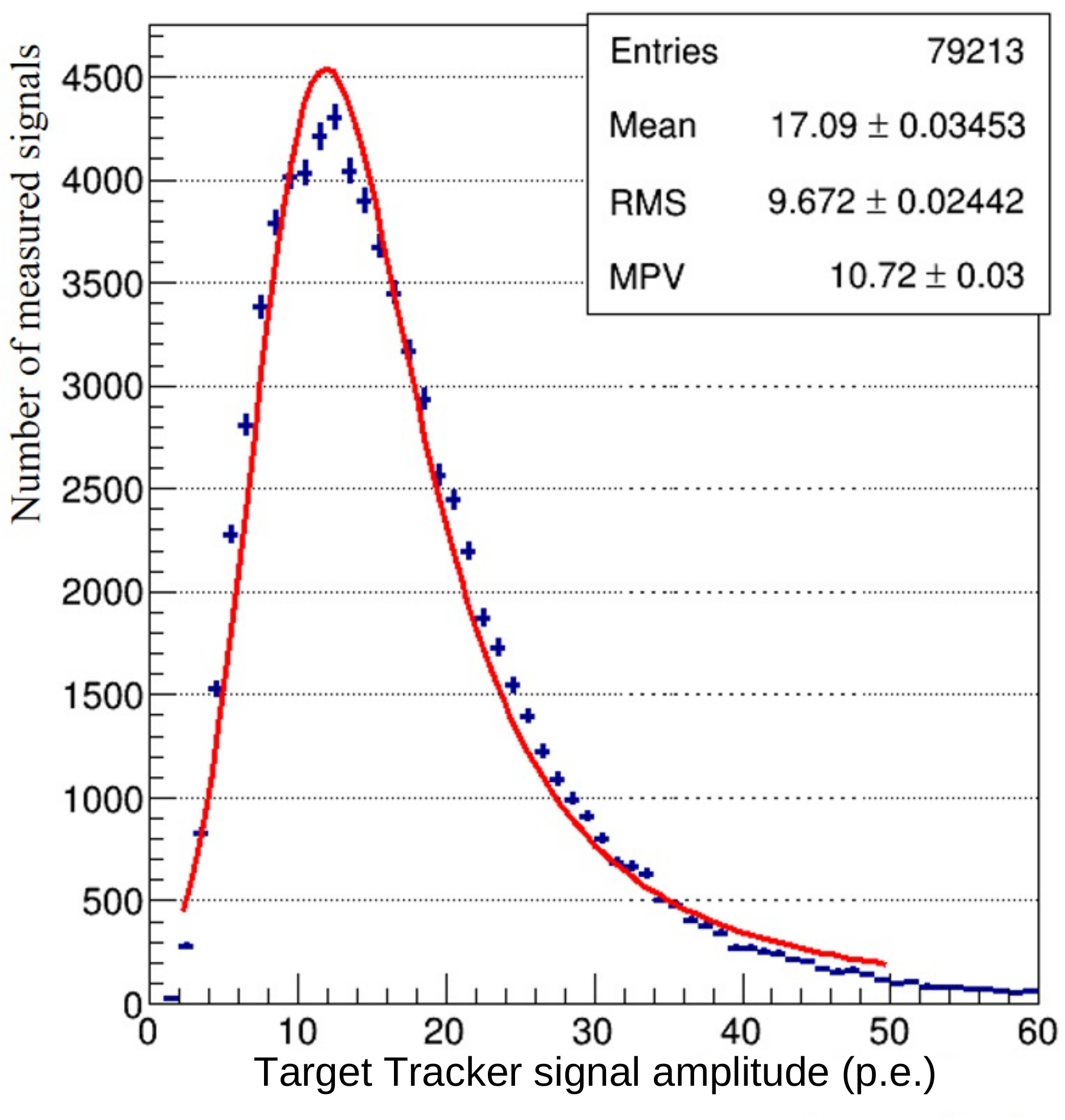}
\caption{Distribution of the OPERA Target Tracker signal amplitudes, the fitted line is a Landau distribution.
  Data used for this distribution was taken during CNGS runs (2008--2012).
  }
\label{ampl}
\end{figure}

From August 2006 until April 2013 electronic cards of the OPERA
experiment were almost continuously registering atmospheric muons along
with muons produced in $\nu_\mu$~CC interactions during the OPERA beam (CERN Neutrinos to Gran Sasso, CNGS) runs.
Over the entire period of the data taking, more than
$10^6$ straight muon tracks were reconstructed in the Target Tracker detector.
Those tracks were used, in particular, to measure the TT
module detection efficiency and to monitor the stability of the TT response to minimum ionising particles.
The TT modules detection efficiency at 1/3~p.e.\@{} threshold was measured
to be equal to ${(98.5\pm 0.3)\%}$ in ``OR'' mode (i.e., when a signal above the
threshold is required from at least one side of a scintillator strip)~\cite{SergeyThesis}.
The ageing of the plastic scintillator was estimated using the time
evolution of the most probable values of the TT response distributions shown in Fig.~\ref{ampl}
obtained for the periods of five years CNGS runs (2008--2012)~\cite{CNGS}.

\begin{sloppypar} 
The characteristic value of the ageing was found to be ${(1.7 \pm  0.2)\%}$ light yield loss per year, which was in agreement with indirect estimates of the plastic scintillator manufacturers ($\sim$20\% over 11.9~years)~\cite{grinev:2003}.
In fact, this evaluation corresponds to the ageing of the entire system also including possible changes of properties of the other TT strips components and materials, like the WLS fibres, glue, etc.
\end{sloppypar}

\begin{figure}[hbt]
\centering
\includegraphics[width=\linewidth]{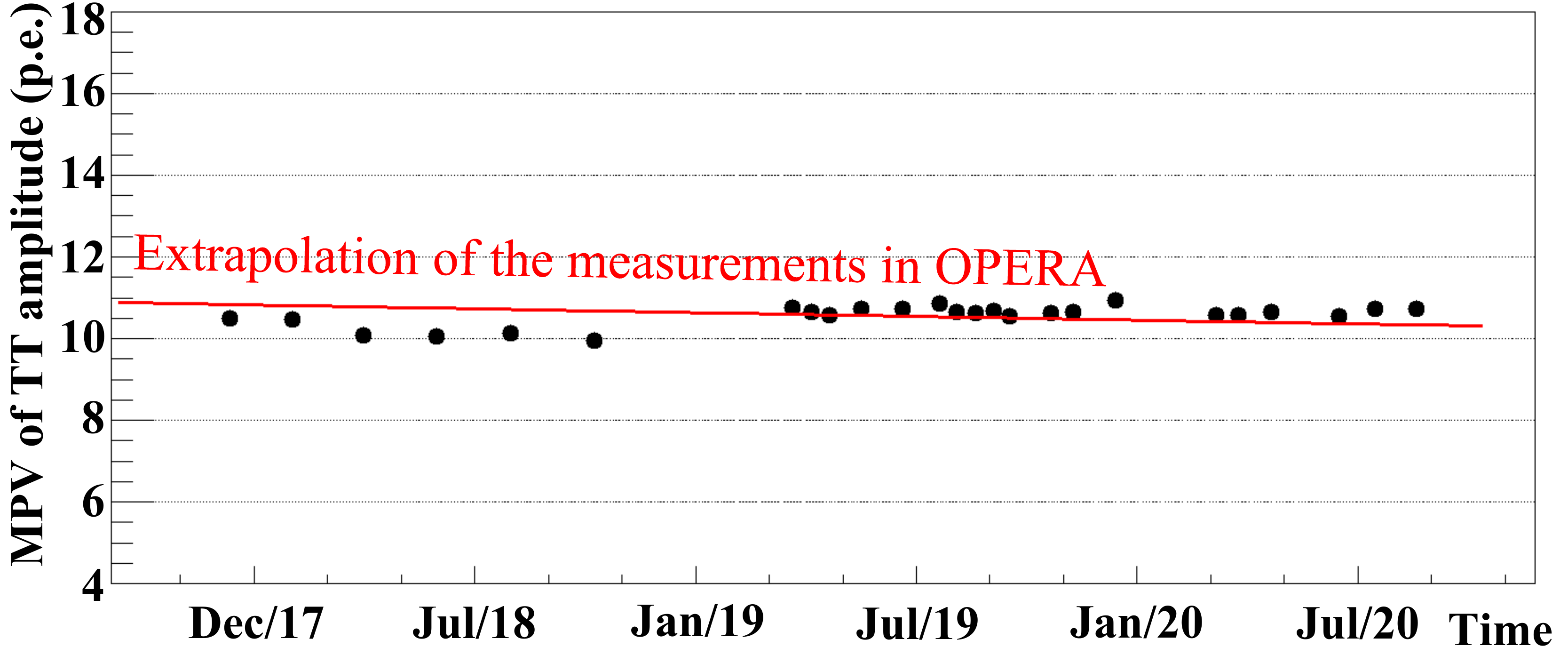}
  \caption{Variation of the most probable value (MPV) of the TT amplitudes with time.
  This data taking period corresponds to when the TT modules were stored in China prior to their assembly in JUNO.
  In early 2019 the TT was moved from a warehouse in Zhongshan to the JUNO site.
  The red line corresponds to the amplitude expected by extrapolating the measured amplitude loss during the CNGS runs.
  }
\label{ageing}
\end{figure}

To follow the ageing of the TT
sensitive materials after dismantling of the OPERA detector and until its use in JUNO, some modules remained equipped with the old OPERA electronics during their transportation to China and subsequent storage.
These modules are used to register atmospheric muons traversing the detector while it remains in storage
prior to it being mounted in JUNO.
This made it possible to still register atmospheric muons with fewer modules.
Although during the period of TT storage, the layout of the containers with the modules have been changed several times, the data collected during each period demonstrated stable average signal amplitude for passing through muons.
It's worth noting that while the change in the layout of the containers can create a systematic change in the
amplitude of the signal for each period, as it was observed when the containers were moved in early 2019,
the relative comparisons within each period would suffer from the same systematic shift and thus
allow us to measure the relative performance deterioration.
The analysis of the collected data shows no additional deterioration of the performance of the plastic scintillator compared to already mentioned ageing (Fig.~\ref{ageing}).

Finally, from their production, about 20~years ago, up to the beginning of the JUNO data taking, about 20\% decrease in the number of p.e.\@{} is expected.
Considering that 6~p.e.\@{} were observed at the middle of each strip (most disfavoured position) after their production, 4.8~p.e.\@{} are expected at the beginning of the JUNO experiment.
With 1/3~p.e.\@{} threshold, the muon detection efficiency remains at comparable level as the initial one.
In case more p.e.\@{} are needed during the JUNO data taking, there is the possibility to introduce optical grease between the fibre opto-coupler and the \mbox{MA-PMT} glass, something which has not been done up to now because it was not needed.
Tests have shown that this will increase by about 15\% the number of detected p.e., compensating a large part of the loss due to ageing.

\section{Summary and Conclusions}
\label{sec:conclusion}

The OPERA experiment Target Tracker has been operated from 2006 to 2013 using the CNGS neutrino beam and atmospheric muons.
After decommissioning, it has been sent to China to be used as Top Tracker (TT) for the neutrino oscillation JUNO experiment.
The available ${6.7 \times 6.7}$~m$^2$ TT walls will be distributed in 3~layers of a ${3 \times 7}$~horizontal grid
above the water Cherenkov detector.
The TT will help to study the background created by the passage of atmospheric muons and hence reduce the systematic error coming from $^9$Li and $^8$He decays mimicking IBD in the JUNO central detector.
Despite not covering the entire surface above the central detector,
the TT will help the other parts of the detector to tune their algorithms to well reconstruct atmospheric muons and veto the affected regions.

For this utilisation, the TT electronics have been modified to cope with the high background induced by the rock radioactivity.
These electronics include for each sensor a Front End Board and a Readout Board.
All 16 sensors of a TT wall are connected to a Concentrator Board whose role is to emit a first trigger level by performing \mbox{x~--~y} coincidences in the same TT wall.
The Concentrator Boards communicate with a Global Trigger Board providing a three-layer TT trigger and with the JUNO data acquisition system.
The trigger rate of the TT after the Global Trigger Board is expected to be around 2~kHz, pushed down to a few Hz
by an offline muon track reconstruction.

In order to validate the whole electronics chain and trigger strategy a prototype has been constructed using the same elements as the Top Tracker.

The TT has been sent to China in 2017 and stored in a hall with a temperature lower than 25~$^\circ$C to keep the plastic scintillator ageing to a reasonable level.
This ageing is expected  to cause a loss of scintillation light of the order of 1\%/year.
Waiting for the TT installation in the underground JUNO site, the scintillator ageing is monitored using atmospheric muons.
Up to now, the observed ageing is compatible with the expectation.
 
The installation of the TT is expected to take place at the end of 2023.
It is evaluated that this installation will last for less than six months.

\section*{Acknowledgments}

\begin{sloppypar} 
We are grateful for the ongoing cooperation from the China General Nuclear Power Group.
We acknowledge the supported of
the Chinese Academy of Sciences,
the National Key R\&D Program of China,
the CAS Center for Excellence in Particle Physics,
Wuyi University,
and the Tsung-Dao Lee Institute of Shanghai Jiao Tong University in China,
the Institut National de Physique Nucl\'eaire et de Physique de Particules (IN2P3) in France,
the Istituto Nazionale di Fisica Nucleare (INFN) in Italy,
the Italian-Chinese collaborative research program MAECI-NSFC,
the Fond de la Recherche Scientifique (F.R.S-FNRS) and FWO under the ``Excellence of Science – EOS'' in Belgium,
the Conselho Nacional de Desenvolvimento Cient\'ifico e Tecnol\`ogico in Brazil,
the Agencia Nacional de Investigacion y Desarrollo in Chile,
the Charles University Research Centre and the Ministry of Education, Youth, and Sports in Czech Republic,
the Deutsche Forschungsgemeinschaft (DFG), the Helmholtz Association, and the Cluster of Excellence PRISMA+ in Germany,
the Joint Institute of Nuclear Research (JINR) and Lomonosov Moscow State University in Russia,
the joint Russian Science Foundation (RSF) and National Natural Science Foundation of China (NSFC) research program,
the MOST and MOE in Taiwan,
the Chulalongkorn University and Suranaree University of Technology in Thailand,
and the University of California at Irvine in USA.
This work, in particular, was supported by
the initiative of excellence IDEX-Unistra (ANR-10-IDEX-0002-02) and EUR-QMat (ANR-17-EURE-0024) from the French national program ``investment for the future'', and
by the Strategic Priority Research Program of the Chinese Academy of Sciences, Grant No. XDA10010300.
\end{sloppypar}

We would like to thank all
private companies which have collaborated with our institutes in
developing and providing materials.
We are particularly grateful to the OPERA Collaboration for the donation of all TT modules to the JUNO Collaboration.
Finally, we would like to thank all technicians non-authors of this paper who have helped us all these last years.

\bibliographystyle{elsarticle-num}
\bibliography{biblio}

\end{document}